\documentclass[namedreferences]{SolarPhysics}
\usepackage[optionalrh]{spr-sola-addons} % For Solar Physics 
\usepackage{graphicx}        % For eps figures, newer & more powerfull
\usepackage{color}           % For color text: \color command
\usepackage{url}             % For breaking URLs easily trough lines
\newcommand{\Deg}{^{\rm o}}

\newcommand{\adv}{    {\it Adv. Spa. Res.}}

\newcommand{\aap}{    {\it Astron. Astrophys.}}

\newcommand{\apj}{    {\it Astrophys. J.}}

\newcommand{\mnras}{  {\it Mon. Not. Roy. Astron. Soc.}}
\newcommand{\nat}{    {\it Nature}}

\newcommand{\solphys}{{\it Solar Phys.}}

\begin{document}

\begin{article}

\begin{opening}

\title{Analysis of MDI High-Degree Mode Frequencies and their Rotational Splittings}

\author{M.C.~\surname{Rabello-Soares}$^{1}$\sep
	S.G.~\surname{Korzennik}$^{2}$\sep
	J.~\surname{Schou}$^{1}$
       }
\runningauthor{M.C. Rabello-Soares {\it et al.}}
\runningtitle{Analysis of MDI High-Degree Mode Frequencies}

   \institute{$^{1}$ {W.W. Hansen Experimental Physics Laboratory,
        Stanford University, 455 Via Palou, Stanford, CA 94305, USA
	email: \url{csoares@sun.stanford.edu} email:\url{schou@sun.stanford.edu}\\
              $^{2}$ Harvard-Smithsonian Center for Astrophysics,
              60 Garden St, Cambridge, MA 02138, USA}
	email: \url{sylvain@cfa.harvard.edu} \\
             }

\begin{abstract}

Here we present a detailed analysis of solar acoustic mode frequencies and their
rotational splittings for modes with degree up to 900. They were obtained by
applying spherical harmonic decomposition
to full-disk solar images observed by the Michelson Doppler Imager
onboard the {\em Solar and Heliospheric Observatory} spacecraft.
Global helioseismology analysis of high-degree modes 
is complicated by the fact that 
the individual modes cannot be isolated, which has limited so far the use of 
high-degree data for structure inversion of the near-surface layers ($r > 0.97 R_{\odot}$).
In this work, we took great care to recover the actual mode characteristics
using a physically motivated model which included a complete leakage matrix.
We included in our analysis 
the following instrumental characteristics: 
the correct instantaneous image scale, the radial and non-radial image distortions, 
the effective position angle of the solar rotation axis and a correction to the Carrington elements.
We also present variations of the mode frequencies caused by the solar activity 
cycle. We have analyzed seven observational periods from 1999 to 2005 and correlated
their frequency shift with four different solar indices.
The frequency shift scaled by the relative mode inertia is a
function of frequency alone and follows a simple power law,
where the exponent obtained for the $p$ modes is twice the value obtained for the $f$ modes.
The different solar indices present the same result.

\end{abstract}
\end{opening}
%-------------------------------------------------

\section{Introduction}
     \label{S-Introduction}

The central frequencies of solar acoustic modes, 
which are obtained using spherical harmonic decomposition,
have been successfully used to determine the solar interior structure
to as close as 21 Mm to the solar surface 
($r < 0.97 R_{\odot}$)
using modes with angular degrees $\ell \le 300$
({\it e.g.}, \opencite{Gough96}).
The inclusion of high-degree modes ({\it i.e.}, up to $\ell = 1000$) has
the potential to improve dramatically the inference of the sound
speed and the adiabatic exponent ($\Gamma_1$) in the outermost 2 to 3\% of the
solar radius, allowing to construct localized kernels as close to 
the solar surface as 1.75 Mm \cite{Rabello-Soares00}.
The effects of the equation of state, through the ionization of hydrogen and helium, 
are felt most strongly in the outer layers of the Sun, making this shallow region of particular interest.
Furthermore, dynamical effects of
convection, and the processes that excite and damp the solar oscillations, are 
predominantly concentrated in this region.
Although the spatial resolution of the modern helioseismic instruments allows us 
to observe oscillation modes up to $\ell$ = 1000 and higher,
only a small fraction of them are currently used ($\ell \le 300$).
Unfortunately, analysis of high-degree data is complicated by the fact that the individual modes
cannot be isolated ({\it e.g.}, \opencite{RKS01}).

The solar structure is not static, but changes over the solar cycle.
It is well known that the mode frequencies change with solar activity.
It seems that the {\color{black} responsible} %causing 
mechanism is restricted to the outer layers of the Sun (\opencite{LW90}),
where the high-degree modes are confined.
However, at the moment, there is no general agreement in the precise physical mechanism that gives
rise to the frequency variation.
It is
likely a product of the change in the subphotospheric small-scale magnetic field strength
with the solar activity cycle ({\it e.g.}, \opencite{Goldreich91}).
Accordingly to \inlinecite{DG04}, the frequency shift is easily explained in terms of a variation
in the turbulent velocities associated with the magnetic field variation rather than the sole direct
effect of the magnetic field itself.
\inlinecite{Li03}, using models
of the structure and evolution of the Sun, 
found that turbulence near the surface of the Sun plays a major role in solar variability, 
and only a model that includes a magnetically modulated turbulent mechanism 
can agree with the observed correlation between the frequency shift and the solar cycle.
In such a dynamic model, 
the evolution of the subsurface layers of the Sun {\color{black} through} %along 
the activity cycle plays an important role.

The frequencies of the global modes give the radius and latitude $(r,\theta)$ part
of the structure.
While, local helioseismic techniques such as ring-diagram analysis \cite{Hill88} 
allow the determination of the three-dimensional structure of the Sun, 
allowing the study of localized areas in the solar surface, such as those in
active regions.
Large variations of the mode frequencies observed in
and near sunspots 
in comparison to magnetically quiet regions 
are well known to be correlated 
with variations in the average surface magnetic field between the corresponding regions.
({\it e.g.}, \opencite{Rajaguru01} and \opencite{RBB07}).
Whether the frequencies are changed directly by the magnetic field or indirectly through an 
associated change in the solar structure ({\color{black} such as} %like 
a pressure change)
is still a matter of debate.
A detailed analysis of the frequency-shift characteristics will hopefully help understand their physical origin.

\inlinecite{Basu04}, using ring-diagram analysis, found that the sound speed is lower 
in the immediate subsurface layers of an active region than of  a magnetically quiet region,
while the opposite is true for depths below about 7 Mm.
However, \inlinecite{Basu02}, using global analysis, have found no observable structural changes 
in the inner layers of the Sun below a depth of 21 Mm associated with the 
magnetic-activity
induced frequency shifts. They were, however, unable to get closer to
the solar surface due to the lack of high-degree modes in their mode set.
High-degree global analysis is important to complete the picture
of the near-surface layers.
Besides, the determination of high-degree frequencies using different methods
allows us to check the results against each other giving
confidence in the results and avoiding systematic errors.
We should point out that, although the high-degree modes have short lifetimes 
(one\,--\,ten %1-10 
hours for $100 \le \ell \le 600$ accordingly to \opencite{Olga07})
propagating only locally, they are averaged over most of the solar surface 
using spherical harmonic decomposition 
(over a relatively long time series)
and thus can still be called global analysis.

In the traditional global helioseismology data-analysis methodology, a time
series of full-disk Doppler solar images is decomposed into spherical harmonic
coefficients, characterized by its degree ($\ell$) and its azimuthal order
($m$).  Each coefficient time series is Fourier transformed, and the order of
the radial wave function ($n$) gets separated in the frequency domain.
However, a spherical harmonic decomposition is not orthonormal over less than
the full sphere -- {\it i.e.}, the solar surface that can be observed from a single
view point-- resulting in what is referred as spatial leakage.  At low and
intermediate degrees, most of these leaks are separated in the frequency domain from
the target mode (except for some $m$ leaks) and individual modes can be identified and fitted. However, at
high degrees, the spatial leaks lie closer in frequency (due to a smaller mode
separation) and, at high frequency, the modes become wider (as the mode lifetimes
get smaller), resulting in
the overlap of the target mode with the spatial leaks that merges individual
peaks into ridges 
(see Figure~1 in \opencite{RKS01}).
The characteristics of the resulting ridge (central frequency, amplitude,
{\it etc}\ldots) do not correspond to those of the underlying target mode. This has
so far hindered the estimation of unbiased mode parameters at high degrees.

To recover the actual mode characteristics, we need a very good estimation of the
relative amplitude of the spatial leaks present in a given $(\ell, m)$ power spectrum,
also known as 
the leakage matrix, which in turn requires a very good knowledge of the
instrumental properties \cite{RKS01}. 
In our previous papers (\opencite{RKS01} and \opencite{KRS04}, hereafter KRS), we described in
detail the large influence of the instrumental properties on the amplitude
of the leaks and as a consequence 
in the determination of unbiased high-degree mode parameters.

In the following, we will first describe the data used in this analysis and the 
ridge-to-mode correction applied to them (Sections~\ref{useddata} and \ref{ridgemodel}).
In Section~\ref{instreff}, we will discuss the influence on the mode parameters of each of the instrumental properties
that were included in the spherical harmonic decomposition of the solar images.
We then analyze in Section~\ref{highdegree} the characteristics of the high-degree mode frequencies and their rotational splittings
obtained in this work.
Finally, in Section~\ref{solarcycle}, we analyze the frequency variation induced by the solar cycle.

\section{Observations and Ridge-Parameter Extraction}
\label{useddata}

The data used in this work consist of full-disk Dopplergrams obtained
at a one-minute cadence by the Michelson Doppler Imager (MDI) 
onboard the {\em Solar and Heliospheric Observatory} (SOHO). 
We have used two distinct sets of data. 
One while MDI was operating in its 4" resolution mode, 
allowing the detection of oscillation modes up to $\ell \approx 1500$,
which we will call from now on the high-$\ell$ data set.
This is the so-called {\em Dynamics Program} observing mode, which is
available every year for two to three months. 
The second one (hereafter referred to as the medium-$\ell$ data set)
using MDI {\em Structure Program} that provides almost continuous coverage year 
{\color{black} round}. %\around. 
In this observing mode,
the original Dopplergrams are convolved, onboard the MDI
instrument, with a Gaussian and subsampled on a $200 \times 200$ grid, thus
reducing the telemetry requirements but also limiting the spatial resolution
to modes with degree $\ell \le 300$.

For the high-$\ell$ set,
we computed the spherical harmonic decomposition of the MDI images for 
modes with $100 \leq \ell \leq 900$.
The resulting time series were Fourier
transformed in small segments (4096 minutes) whose spectra were averaged
to produce an averaged power spectrum%\footnote{The number of averaged ...
with a low but adequate frequency resolution to 
fit the ridge while reducing the realization noise.
{\color{black}The number of averaged spectra varies with the year, from 12 segments in 2003, to 30 in 2001.}
Most of the known instrumental effects relevant to the high-degree
analysis were included in the spatial decomposition and they will be described in
Section~\ref{instreff}.
The peaks in each $(\ell,m)$ spectrum were then fitted using an asymmetric Lorentzian profile 
with an additive background term, given by ten to the power of a {\color{black}second-}degree polynomial in frequency
(Equation~5 in KRS).
Since the number of segments used in the average
of each $(\ell,m)$ spectrum is large enough, %(see footnote 1), 
its $\chi^2$ distribution 
can be approximate by a Gaussian distribution and
a least-square fitting was used. 
The fitted Lorentzian profile is characterized by the following parameters: 
frequency ($\nu_{n,\ell,m}$), amplitude ($A_{n,\ell,m}$) , width ($\gamma_{n,\ell,m}$) and, asymmetry ($\alpha_{n,\ell,m}$).
The asymmetric profile used is equivalent to the one defined by \inlinecite{Nigam}
where their asymmetric parameter is equal to $\alpha / (2 - \alpha)$.
The frequency splittings were parametrized in terms of 
Clebsch-Gordan coefficients up to $a_6$ \cite{Ritz91}. 
The number of modes analysed was reduced 
without loss of information to the work described in this paper
and hence {\color{black}was} easy to handle.
The fitting was carried out only every tenth $\ell$ and
only for some 50 equally spaced $m$ values at each $\ell$.
The central frequency, {\it i.e.} the frequency free of splitting effects,
is taken to be the frequency given by %making 
$m = 0$ in the splitting parametrization.
Since the even splitting coefficients are zero on average, 
it is the same as the mean frequency (averaged over $m$).
Notice that the fitted parameters are ridge parameters and do not correspond to 
the associated mode parameters as discussed in the introduction.
In this study, high-$\ell$ time series available from 1999 until 2005
were used and their properties are listed in Table~\ref{table_epochs}.

\begin{table}
\caption{Details of the analyzed time series. The corresponding relative mean values 
of solar UV spectral irradiance and their standard deviation
are also listed as an indication of the solar activity level.}
\label{table_epochs}
\begin{tabular}{ccclc}

\hline
{Year}  & {Starting Date} & {Duration} & {\,\,\,\,\,\,\,\,Solar Index} &  {Starting Date}  \\
 	& high-$\ell$ set    & high-$\ell$ set  & {rel.~to max. (in \%)} &  medium-$\ell$ set      \\
\hline
%--------------------------------------------------------------------------------------
 1999 & Mar. 13 & 77 days &  \,\,\,\,\,\,\,\,\,\,40$\,\pm\,$13         &  Feb. 03 \\
 2000 & May  27 & 45 days &  \,\,\,\,\,\,\,\,\,\,69$\,\pm\,$11         &  Apr. 10 \\
 2001 & Feb. 28 & 90 days &  \,\,\,\,\,\,\,\,\,\,61$\,\pm\,$14         &  Jan. 23 \\
 2002 & Feb. 23 & 72 days &  \,\,\,\,\,\,\,\,\,\,80$\,\pm\,$8          &  Mar. 31 \\
 2003 & Oct. 18 & 38 days &  \,\,\,\,\,\,\,\,\,\,51$\,\pm\,$17         &  Oct. 28 \\
 2004 & Jul. 04 & 65 days &  \,\,\,\,\,\,\,\,\,\,36$\,\pm\,$11         &  Aug. 11 \\
 2005 & Jun. 25 & 67 days &  \,\,\,\,\,\,\,\,\,\,30$\,\pm\,$9          &  May  26 \\
\hline
\end{tabular}
\end{table}
%--------------------------------------------------------------------------------------

\begin{figure}
\centerline{\includegraphics[width=0.8\textwidth,clip=]{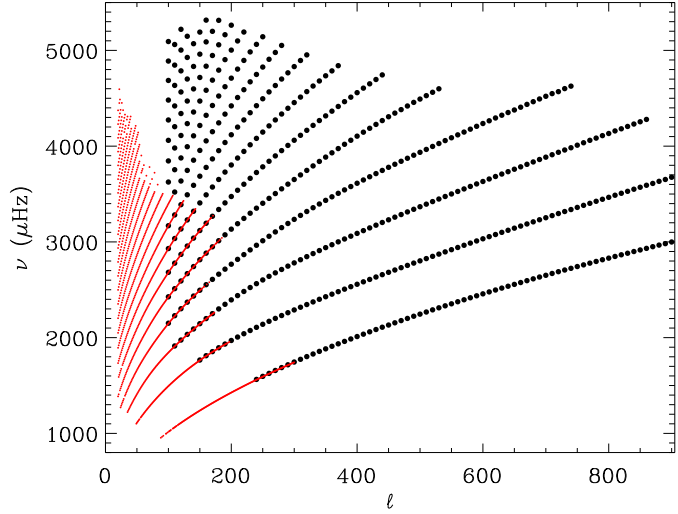}
	   }
   \caption{ Coverage, in an $\ell-\nu$ diagram, of the medium-$\ell$
(red) and high-$\ell$ (black) mode parameters
for 2005. The coverage is very similar for all epochs.
}
\label{fig:lnu}
\end{figure}

The focus of this work are modes with $\ell > 300$ and therefore
it is centered on the analysis of the high-$\ell$ set. We used however the
results obtained by one of us analyzing the medium-$\ell$ time series
\cite{Schou99}
to compare with and complement our analysis.
Using 72-day long time intervals, the central frequency and splitting coefficients
for a given $(n,\ell)$ mode were determined directly by
fitting symmetric Lorentzian profiles to its power spectra \cite{Schou99}.
Every $p$ mode up to $\ell \approx 200$ (up to $\ell \approx 300$ for the $f$ modes)
and every $m$ were fitted. 
The 72-day time series that best overlap in time with the high-$\ell$ time series were selected
and are listed in Table~\ref{table_epochs}.
Only modes with $\ell \geq 20$ were used.
The mode coverage of both analyses is illustrated in Figure~\ref{fig:lnu}.

%---------------------------------------------------------
%---------------------------------------------------------

In the medium-$\ell$ analysis, 
the instrumental effects described in Section~\ref{instreff} were not taken into account,
and neither were the
distortion of the eigenfunctions by the solar differential rotation 
or the horizontal component of the oscillation (both described in Section~\ref{ridgemodel}).
However, in the frequency and degree ranges of the medium-$\ell$ analysis,
the spatial leaks are well separated from the target mode and
individual modes can be identified and fitted,
making the above-mentioned effects
not as crucial as {\color{black}they are} %it is 
for the high-$\ell$ analysis.
Another relevant difference between the medium- and high-$\ell$ analysis is that
the medium-$\ell$ power spectra were fitted using symmetric profiles,
which is well known to lead to systematic errors in the frequency measurements
({\it e.g.} \opencite{Toutain98}; \opencite{Basu00}).  %In a recent paper, \inlinecite{Larson08} re-analyzing the medium-$\ell$ time series
\inlinecite{Larson08} {\color{black}have recently re-analyzed} the medium-$\ell$ time series
{\color{black}and}
reported that several of these 
physical effects 
result in highly significant 
changes in the mode parameters.
Their Figure~1 shows the total correction 
to be applied to the medium-$\ell$ frequency and splitting coefficients, $a_1$ and $a_2$,
used here.  %And 
Their Figure~2 illustrates the
frequency changes due to each of these effects. %\footnote{
The first two panels correspond to the distortion of the eigenfunctions
by the solar differential rotation and the horizontal component of the leakage matrix 
respectively.
The third, forth, and fifth panels 
correspond to the instrumental effects described here in Sections~\ref{sec:imagescale}, \ref{sec:radial}, and
\ref{sec:nonradial} respectively. The panels at the bottom show the difference obtained between
fitting symmetric and asymmetric Lorentzians.
Although these corrections are significant,  
they correspond to very small variations in the results presented in this paper
and do not affect our conclusions,
as it will be discussed later (Section~\ref{highdegree}).

\section{Ridge-to-Mode Correction}
\label{ridgemodel}

Our methodology to recover the mode characteristics 
from the ridges observed at high-degree and high-frequency power spectra
consists in generating and
fitting a sophisticated model of the underlying modes that contribute to the
ridge power distribution and deduce the offset ($\Delta^\circ$) between the ridge properties
and the target mode \cite{Korzennik98}.
A synthetic power spectrum is computed for each $(\ell,m)$ mode 
consisting of several asymmetric Lorentzians for each $n$: one for the target mode $(\ell, m)$ and 
one for each spatial leak $(\ell', m')$ with a relative amplitude given by the leakage
matrix.
It has the same frequency resolution as the high-$\ell$ set power spectra and
it is generated for the same set of $(\ell, m)$ modes.

We used the complete leakage matrix ({\it i.e.}, radial and horizontal components)
where the horizontal-to-vertical displacement ratio is given by
\inlinecite{CD03}:
$G M_\odot L / (R^3_\odot \omega^2_{n, \ell})$,
where $G$ is the gravitational constant, $M_\odot$ is the solar mass, $R_\odot$ is the
solar radius, $\omega$ is the cyclic frequency ($\omega = 2\pi\nu$), and $L^2 = \ell(\ell+1)$.
For each $(\ell, m)$ synthetic power spectrum, we have taken into account the contribution
of the spurious modes $(\ell', m')$ that obey the following expressions:
$|m' - m| \leq 10$ and $|\ell' - \ell| \leq \ell_d$, where $\ell_d=12$ for
$\ell \leq 600$ and $\ell_d$ is equal to $0.02 \, \ell$ rounded to the closest integer
for $\ell > 600$.
Spurious modes 
with $(\ell', m')$ values that differ 
from those of the power spectrum $(\ell, m)$
by more than the amount specified in the equations above
have a very small relative amplitude 
and their contribution to the synthetic power spectrum profile is negligible, 
{\it i.e.},
their inclusion or not {\color{black}does} not affect the fitted parameters of the peaks in the power spectrum
\cite{RKS05}.
We also included in our model the distortion of the eigenfunctions by the solar
differential rotation as described in \inlinecite{Woodard89},
using the values given by \inlinecite{Schou98} for the solar rotation.
The coefficients in this superposition become negligible when 
$|\ell - \ell'| \geq \ell_c$ where $\ell_c = 10$ for $\ell \leq 400$ and 
the next even integer to $3 + 0.02 \, \ell$ for $\ell > 400$ \cite{RKS05}.

The profile resulting from the overlap of several profiles of nearby spatial leaks is reasonably well
modeled by a single profile when the ratio of the mode 
width to their separation (given by $\partial\nu/\partial\ell$) becomes large. 
The difference is smaller than the rms of the observed residuals to the fit, 
5\%\,--\,7\% at all degrees, which are dominated by the realization noise (KRS).
The synthetic power spectrum is then fitted 
following the methodology applied to the high-$\ell$ set (Section~\ref{useddata}), 
providing the ridge central frequency $\nu_{n,\ell}^{{\tiny\mbox{model}}}$ and 
its splittings $a_{i_{n,\ell}}^{{\tiny\mbox{model}}}$
($i$ = 1,6).
The offset is given by the difference between the 
modeled ridge and the target mode parameter given by the input value
of the parameter used to generate the synthetic power spectra.
For a given mode $(n,\ell)$, the offsets in central frequency can be then written as:
\begin{equation}
\Delta^\circ\nu_{n,\ell} = \nu^{{\tiny\mbox{model}}}_{n,\ell} - \nu^{{\tiny\mbox{input}}}_{n,\ell}.
\label{eq:offset}
\end{equation}
The offsets in each of the fitted $a$-coefficients are obtained in the same manner.
Realistic input values based on observed mode parameters were used (as described in KRS).
The input linewidth is given by
the square root of $\gamma^2_{{\tiny\mbox{mode}}} + \gamma^2_{{\tiny\mbox{WF}}}$,
where 
$\gamma_{{\tiny\mbox{mode}}}$ is an estimate of the mode linewidth and
$\gamma_{{\tiny\mbox{WF}}}$ is the width of the window function
in our case given by the length of the high-$\ell$ set time string ({\it i.e.}, 4096 minutes).
This expression was obtained by \inlinecite{Korzennik90}
assuming that the observed power spectrum is a convolution of 
the ``true'' power spectrum 
({\it i.e.},  the one that would have been obtained with a
infinite time series), 
with the power spectrum of the window function
both represented by Gaussian profiles,
which is adequate for the purpose {\color{black}at hand}. %in view.

If the leakage matrix is correct and complete, our simulations should adequately estimate the
parameter offsets ($\Delta^\circ$) and we would be able to 
obtain corrected mode parameters from the observed ridge. 
Thus in the case of the central frequency, we would have:
\begin{equation}
 \nu^{{\tiny\mbox{corrected}}}_{n,\ell} = \nu^{{\tiny\mbox{observed}}}_{n, \ell} - \Delta^\circ\nu_{n, \ell}.
\label{eq:offset2}
\end{equation}
The corrected mode $a$-coefficients are obtained as in Equation~(\ref{eq:offset2}).

\begin{figure}
\centerline{\includegraphics[width=0.8\textwidth,clip=]{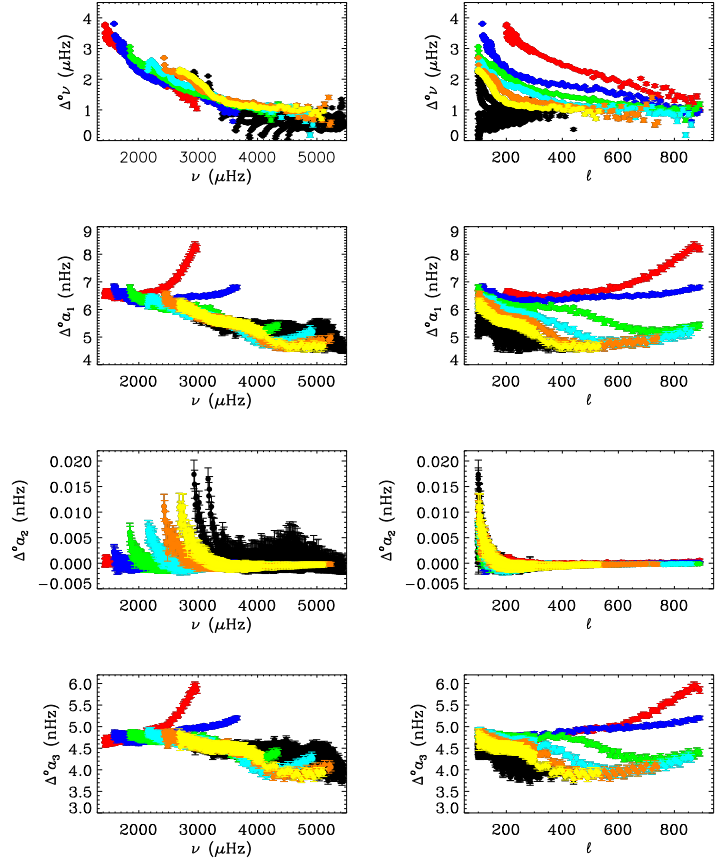}
	   }
\caption{
Estimated offsets for the central frequency ($\nu$) and splitting coefficients ($a_1$, $a_2$, and $a_3$).
The error bars correspond to the uncertainties when fitting the synthetic power spectra.
Modes with $n \le 5$ are in color {\color{black} with %where 
the $f$ modes in red}.
}
\label{fig:offset}
\end{figure}

Figure~\ref{fig:offset} shows the estimated offsets for the central frequency and
the splitting coefficients $a_1$, $a_2$, and $a_3$.
They are {\color{black} primarily} %mainly 
a function of frequency.
Except for $a_2$, the offsets are quite large in comparison with the observed fitting uncertainties
of the corresponding parameter which, for the central frequency, $a_1$ and $a_3$ coefficients,
are usually smaller than 0.5 $\mu$Hz, 1 nHz, and 0.8 nHz respectively.
The width and amplitude offsets will be described in a future paper.
Accordingly to \inlinecite{Korzennik98}, the asymmetry of the ridge seems to be the same
as the asymmetry of the underlying mode (see also KRS).

%.............................................................................
%.............................................................................

Another methodology to recover the mode characteristics from the observed
ridges is to fit a sum of Lorentzians or asymmetric profiles to a given ridge:
one for the target mode and one for each spatial leak
that has a significant amplitude (\opencite{Reiter02};
\opencite{Reiter03}; \opencite{Reiter04}).
Such an approach is likely to be particularly appropriate
in the transition region between well-resolved modes and ridges,
where the spatial leaks start to overlap with
the target mode but do not yet blend fully into a ridge.
The reasoning behind the method used here is that,
when the spatial leaks are completely blended into a ridge,
there is not enough information 
in the power spectra to justify modeling the individual spatial leaks
as part of the fitting of a given observed spectrum.
The difference in the profile of a sum of overlapped asymmetric profiles
and a single asymmetric profile is much smaller than the observed fitting
residuals.
This method is significantly less computationally demanding since the
profiles fitted are much simpler.
As a result we can more easily
check its {\color{black}reliability} %and robustness, 
by using different leakage matrices or
time series produced with a different spatial decomposition, and
quantitatively validating the corrections; as described in
Section~\ref{instreff} and KRS.
This validation step ensures the reliability of the method and
allow us to estimate a quantitative upper limit on any residual
bias, a crucial step in the intricate high-$\ell$ analysis.
Note that the method described by Reiter {\it et al.} also relies on a very good
estimation of the
leakage matrix to obtain unbiased mode parameters. Indeed, the same
leakage matrices have been used for both methods and it is likely
that both the random and systematic errors will be quite similar
where the modes are fully blended into ridges.

\section{Influence of Instrumental Properties on the High-Degree Power Spectra}
\label{instreff}

In the determination of unbiased high-degree mode parameters,
it is necessary to have a good knowledge of the properties of the 
instrument used to collect the data.
The instrumental characteristics must be 
taken into account either in the image spatial decomposition or in the leakage matrix calculation
to obtain a correct estimation of the amplitude of the spatial leaks.
In a continuing effort to infer unbiased estimates of high-degree mode parameters
using the high-resolution observations from the MDI {\em Dynamics Program},
the following instrumental properties were included in the spatial 
decomposition of the high-$\ell$ data set one at a time
and their effect on the observed power spectra analyzed:	
({\it i}) the correct instantaneous image scale, ({\it ii}) the radial
image distortion, ({\it iii}) the non-radial image distortion,
({\it iv}) the effective $P$ angle, and ({\it v}) a correction to the
Carrington elements.
Once a given instrumental property is analyzed, it is incorporated in the analysis
from then on in the paper. 
For example, in the analysis of the radial image distortion (Section~\ref{sec:radial}),
the correct image scale (Section~\ref{sec:imagescale}) was used and, 
in the non-radial image distortion analysis (Section~\ref{sec:nonradial}), 
the correct image scale and the radial image distortion 
were included.

\subsection{Image Scale}
\label{sec:imagescale}

Variations in the amount of defocus of the instrument have a direct influence in the image scale
at the detector\footnote{{\color{black}Image scale is the ratio between the size of the solar image at the detector and its actual size.}}.
Although the MDI instrument has been very stable over the more than 11 years
that it has been observing the Sun, continuous
exposure to solar radiation has increased the instrument front window
absorption resulting in a continuous small increase of the instrument defocus. 
Moreover the change of the front-window temperature due to the
satellite orbit around the Sun also adds a small annual variation
in the image defocus. The instrument has however an adjustable focus with nine possible
positions
chosen to best suit various science needs, resulting unfortunately in
abrupt jumps in the image defocus every time that a new position is chosen and 
which are responsible for the largest variations 
(see Figure~5 in \opencite{RKS01}).
The average size change (at the solar limb) per focus step is 0.529$\pm$0.002 pixels \cite{Kuhn04}.
Due to these different time-varying focus variation,
the image scale must be 
continuously
estimated and the correct value used in the spatial decomposition.
The image scale is obtained by measuring the observed image radius, which is defined as
the inflection point in the radial limb-darkening function. %\footnote{
{\color{black}
The
\url{FNDLMB} routine} in the GONG Reduction and Analysis Software Package (\url{GRASP}) was used.
It is available from the National Solar Observatory, Tucson, AZ, U.S.A..

\begin{figure}
\centerline{\includegraphics[width=0.8\textwidth,clip=]{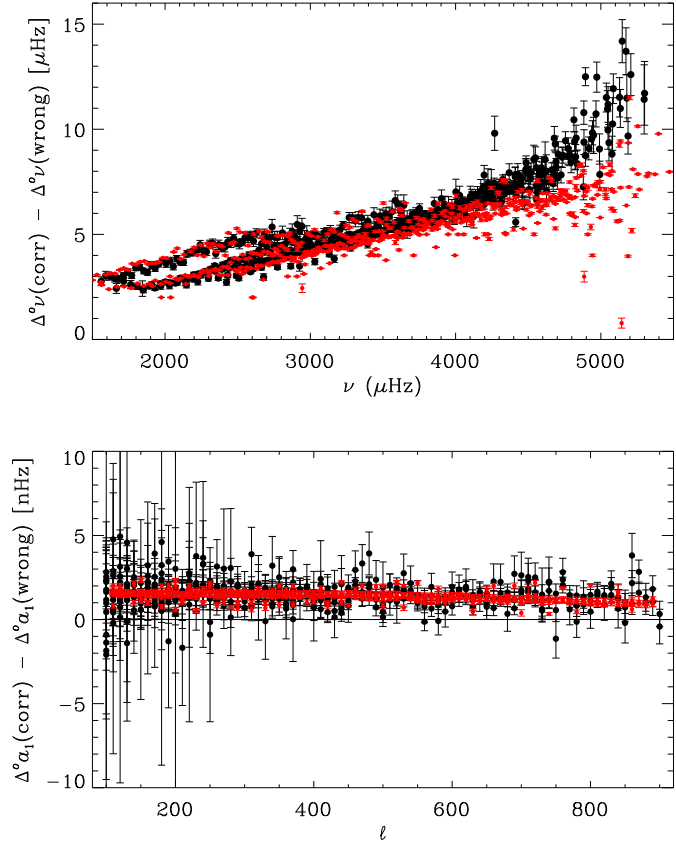}
	   }
\caption{
Difference in the frequency and the $a_1$-coefficient offsets using,
in the spatial decomposition for the 1999 high-$\ell$ set,
the correct time-varying image scale (``corr'') and 
a constant value, which is 0.27\% larger on average than the actual values (``wrong'').
The errors are given by the fitting uncertainties.
{\color{black}
The estimation of this effect 
when it is included in the leakage matrix calculation instead
(see text for details) is shown in red.
}
}
\label{fig:imagescale}
\end{figure}

To show the influence of the image scale in the high-$\ell$ mode parameters, we compared the
observed ridge parameters obtained using two different spatial analyses of the 1999 time series.
In one of the {\color{black}analyses}, the time-varying image scale was obtained for the actual observational period
({\it i.e.}, the correct image scale) and used in the spatial decomposition.
In the other one, a constant value obtained from observations taken in early 1996 (at the beginning of the mission),
{\it i.e.}, the wrong value for the 1999 time series, was used. It is
0.27\% larger on average than the actual 1999 time-varying image scale.
{\color{black}A variation in the observed ridge parameter due to a change in the data
analysis corresponds to an identical variation in its offset
($\Delta^\circ$) in order to obtain the correct mode parameter (see Equation~\ref{eq:offset2}). 
Figure~\ref{fig:imagescale} shows the corresponding offset variation.}
The frequency offsets obtained using the correct image scale
are systematically larger than the ones using a slightly larger image scale 
and their difference increases with frequency (Figure~\ref{fig:imagescale}).
{\color{black}There is no indication of a degree dependence.}
The variation in the frequency offset is larger for the $f$ modes than for $p$ modes
(respectively upper and lower branches seen at frequencies smaller than 3 mHz
in Figure~\ref{fig:imagescale}). 
This suggests that the image scale
correction has a strong effect on the horizontal component of the leakage matrix.
The $a_1$ coefficients obtained using the correct image scale
are systematically larger by 
$1.44 \pm 0.02$ nHz 
than using an image scale 0.27\% larger on average.
{\color{black}
Their difference is plotted in Figure~\ref{fig:imagescale} (bottom panel) arbitrarily against degree
instead of frequency.}
The effect on the other parameters is very small and 
their mean difference is listed in Table~\ref{variation}.
The 2000 data set was used to calculate the values in Table~\ref{variation},
except for the variations due to the effect of the image scale where the 1999 epoch was used. 
The values in the table should not depend significantly on the observational period used
since we are comparing variations in the analysis of the same time series.
The variations in the frequency offset (Figure~\ref{fig:imagescale} top) are quite large
in comparison with their absolute values shown in Figure~\ref{fig:offset}.
The theoretical offset estimation in Figure~\ref{fig:offset} corresponds to the 
case where the correct values for all
instrumental effects were taken into account in the spherical
harmonic decomposition.

{\color{black}
The red points in Figure~\ref{fig:imagescale} were calculated taking into account the 
image scale error in the leakage matrix calculation instead of in the image spatial decomposition.
The leakage matrix is calculated assuming that a constant and 0.27\% larger image scale than the actual value
was used in the image spatial decomposition.
The variation in the parameter offsets estimated from the synthetic power spectra using Equation~(\ref{eq:offset})
matchs very well the observations in most cases.}
For modes above 4 mHz, the frequency changes were underestimated by 
this method.
This could be because
we did not take into account the image-scale {\color{black}temporal} %time 
variation, only the average difference for the
entire observing period, while the spatial decomposition was carried out using a
instantaneous image scale estimation.

%.............................................................................
%.............................................................................

Note that the smearing of the image represented by the point-spread function (PSF)
is not taken into account nor its variation with the amount of defocus.
Our preliminary analysis indicated that 
including a simple model of the azimuthally averaged estimation of the PSF in the leakage matrix calculation 
has a very small effect on the offsets. It affected mostly the frequency offset and only by less
than 0.2 $\mu$Hz \cite{RKS06}.
{\color{black}Recently} %Lately 
we found that the observed ridge frequencies obtained for the
observational periods where %the 
MDI was set to a large defocus, {\it i.e.}, 1996 to 1998, 
are larger than the ones obtained for the other periods where the instrument was nearly {\color{black}in} focus,
after correcting for the solar-cycle variation; their maximum difference 
is of the order of a few $\mu$Hz \cite{RKS08}.
A possible explanation is that an azimuthally averaged estimation of the PSF is not a good
approximation to the true PSF of the instrument,
which is known to depend on the azimuth angle \cite{SchouBogart}
with a phase that changes with focus position ({\it e.g.}, KRS).
Unfortunately, there is not a good {\color{black}estimate} of the PSF for the MDI 
{\em Dynamics Program} at the moment.  %As a future work, we plan 
We plan to use an approximation to the
azimuthally-varying PSF and analyze its influence in the mode parameter determination
{\color{black} in the future.}

%-----------------------------------------------------
\begin{table}[h]
\caption{Mean differences in the observed ridge parameters in the sense improved minus unimproved.
The weighted average and its standard error {\color{black}were} calculated over all 
observed {\color{black}modes (500 in all cases).
The mode range is shown in Figure~\ref{fig:lnu} (black circles).}
The weight of each determination was taken as inversely proportional to the square of the uncertainty.
The uncertainties of the central frequency and the splitting coefficients are given by 
the uncertainties in the Clebsch-Gordan expansion.
The uncertainties in the width, amplitude, and asymmetry are given by the standard deviation 
of the $m$-average.
The first line (``Uncertainty'') gives the minimum and maximum values of the uncertainties,
which are very similar in all the analysis presented in the Table. 
The other lines, from top to bottom, are the differences in the image scale, radial image distortion, 
non-radial image distortion, $P$-angle and Carrington elements respectively.
The amplitude values are given in arbitrary units and the asymmetry parameter is dimensionless.
}
\label{variation}
\begin{tabular}{ccccc}     % define the column alignment
                           % l: left, c: center, r: right
  \hline                   % horizontal line
Parameter  & Frequency & $a_1$ & $a_2 \times10^5$        &        $a_3$ \\
           & ($\mu$Hz) & (nHz) & (nHz) &  (nHz) \\
  \hline
Uncertainty & 0.05 -- 1.4 &  0.07 -- 2.0 & 70 -- 20000 & 0.2 -- 8 \\
Scale & $4.54 \pm 0.06^a$ &$ 1.44 \pm   0.02$ &$-25\pm9$ & $0.03 \pm 0.02$ \\
Radial & $0.95 \pm     0.01^a$ & $0.42 \pm     0.02$ & $-95\pm7$ & $0.03 \pm 0.02$ \\
Non-rad. &   $0.133 \pm 0.008$ &  $-0.11 \pm  0.02$ &  $240\pm10$ &  $0.03 \pm  0.01$ \\
P angle &  $-0.0012 \pm    0.006$ &   $0.04 \pm     0.02$ &  $-0.5\pm8$    &  $0.025 \pm     0.016$ \\
Carr. el.&  $0.00020 \pm    0.005$  & $-0.018 \pm     0.02$    & $-1\pm6$   & $-0.013 \pm     0.01$ \\
  \hline
\end{tabular}

\begin{tabular}{cccc}     % define the column alignment
  \hline
Parameter  & Width ($\mu$Hz) & Amplitude & Asymmetry \\
  \hline
Uncertainty & 0.3 -- 6 & 50 -- $10^5$ & 0.003 -- 0.12 \\
Scale & $0.047 \pm   0.007$ & $-4\pm2$ & $(-1\pm2)\times10^{-4}$ \\
Radial & $-0.018 \pm    0.005^a$ &  $0.4\pm1$ &  $(-7.5\pm0.5)\times10^{-4}$ \\
Non-rad. & $-0.044 \pm 0.005$ &  $2\pm1$ &  $(1\pm0.4)\times10^{-4}$ \\
P angle & $-0.00026 \pm    0.005$ &  $-2\pm2$ &  $(9\pm4)\times10^{-7}$ \\
Carr. el. & $-0.0050 \pm 0.004$  & $0.9\pm1$ & $(-0.4\pm4)\times10^{-5}$ \\
  \hline
\end{tabular}

$^a$ In these cases, the difference depends on the frequency (see text).

\end{table}
%-----------------------------------------------------

\subsection{Radial Image Distortion}
\label{sec:radial}

The ray-trace model of the MDI optical configuration predicts a radial distortion 
($\Delta r$) which depends on the distance from the CCD center ($r$):  
\begin{equation}
  \frac{\Delta r}{r} = b \; (r^2 - <\!\! r \!\!>^2),
\label{eq:raddist}
\end{equation}
where $b = 7 \times 10^{-9}$ pixels$^{-2}$ and $<\!\! r \!\!>$ is the observed image mean radius \cite{Kuhn04}.
The distortion causes the apparent solar radius to be larger by $\approx$0.17\% ($\approx$0.8 pixels or 17 $\mu$m).
Thus, 
the second term in Equation~(\ref{eq:raddist}) was added to ensure that the distorted and undistorted images have the same mean radius.

\begin{figure}
\centerline{\includegraphics[width=0.8\textwidth,clip=]{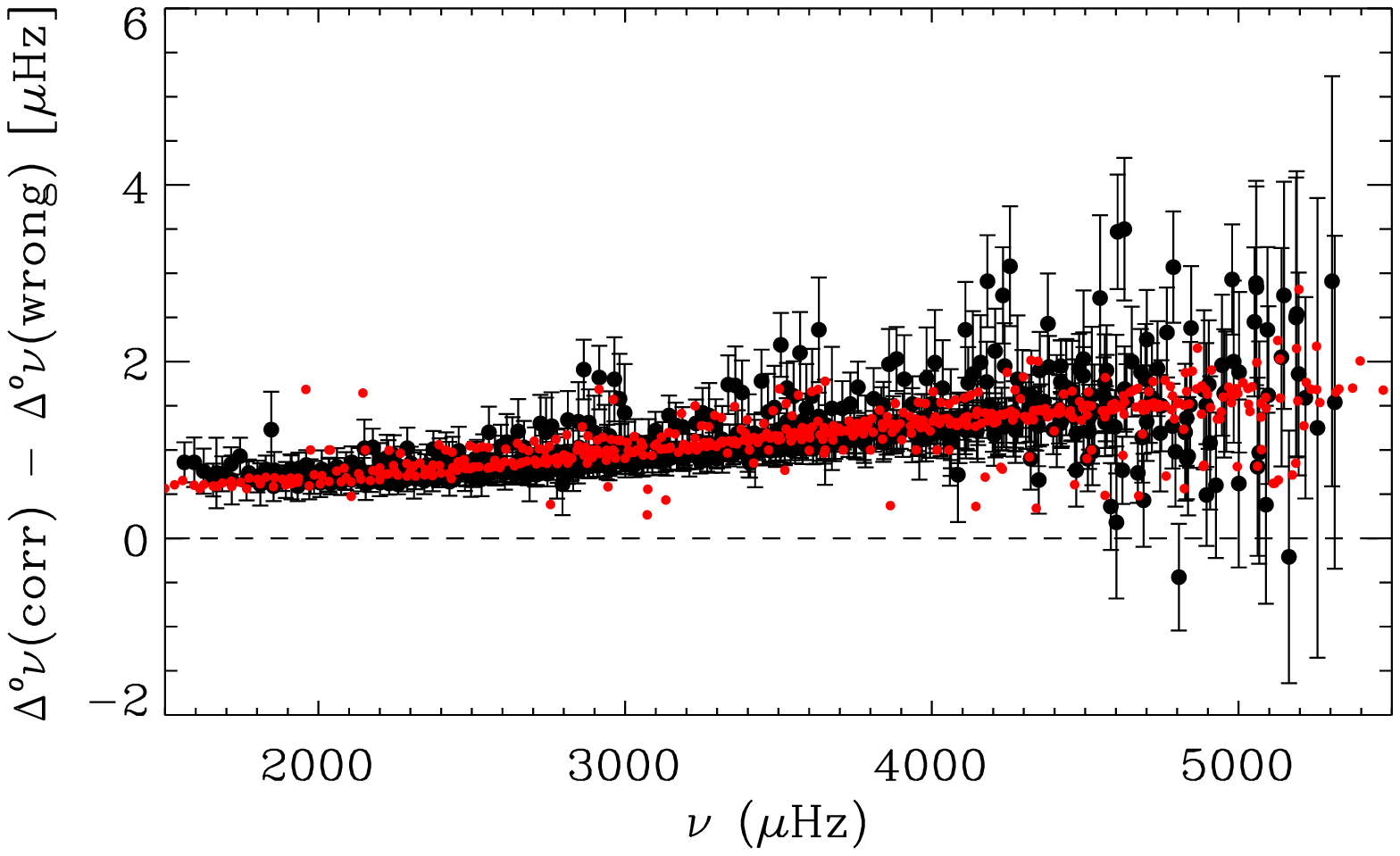}
	   }
\caption{Difference in the frequency offset obtained when including or not the radial 
distortion in the spherical harmonic decomposition using the 2000 high-$\ell$ data set.
The errors are given by the fitting uncertainties.
The estimation of this effect using our model is shown in red.
}
\label{fig:radist}
\end{figure}

{\color{black}
As in Figure~\ref{fig:imagescale},
Figure~\ref{fig:radist} shows
the offset variation corresponding to the difference in the observed ridge frequencies when including this distortion in the spatial decomposition.
}
Similarly to the image scale, the frequency offset increases with frequency.
The similarity is expected, since the
radial distortion changes the image scale by an amount that is a function
of the distance from the CCD center $r$ (Equation~\ref{eq:raddist}).
The difference in the ridge width also increases with frequency
from $-0.064 \pm 0.004$ $\mu$Hz at $\nu < 2.5$ mHz to $0.28 \pm 0.04$ $\mu$Hz at $\nu > 4.5$ mHz,
in the sense including minus not including the distortion,
but it is very small in comparison with the fitting uncertainties and barely significant.
The effect on the other parameters is small and it is listed in Table~\ref{variation}.
The ridge modeling of this effect, introducing the radial distortion in the leakage matrix,
agrees well with the observations (red circles in Figure~\ref{fig:radist}).

%.............................................................................
%.............................................................................

\subsection{Non-Radial Image Distortion}
\label{sec:nonradial}

Solar images as observed by MDI have a nearly elliptical shape with a
difference between the semi-major and the semi-minor axis of about 0.6 pixels. 
This is consistent with a small tilt ($\approx$2$\Deg$) 
in the detector around an axis that is rotated 56$\Deg$ from
the detector's horizontal $x$-axis and it
introduces a non-radial image distortion.
In KRS, we estimated this distortion using different observational methods.
Unfortunately, the distortion varies by as much as 35\% depending on the data used,
with the correspondent CCD tilt varying from 1.71$\Deg$ to 2.6$\Deg$.
\inlinecite{Kuhn04} also estimated the non-radial distortion using yet another observational 
method -- the Mercury transit across the Sun on 7 May 2003 -- and found a 3.3$\Deg$ tilt.
Although their equations to estimate the distortion have the same general form as ours (KRS), 
there are other differences besides the tilt angle between the two calculations.
Their estimation of the distortion varies by 14\% in relation to our estimation using a 2.6$\Deg$ tilt.
These discrepancies in the estimated distortion
using different observational methods
might be attributed to a number of reasons such as
the inaccuracy of our simple model for the non-circular shape of the solar images or
the influence of an optical aberration (an asymmetric PSF, for example).

To analyze the effect of the non-radial distortion on the power spectra, we included
in the spherical harmonic decomposition our estimation that has the largest tilt angle
(2.6$\Deg$),
which corresponds to the distortion that better reproduces the solar limb shape.
Although, to the moment, we were unable to determine precisely the non-radial distortion pattern,
our estimation provides an improvement to the analysis and it 
will be incorporated from then on in the paper.
Fortunately, {\color{black}it has an overall small effect on the observed spectra,} as can
be inferred from Table~\ref{variation}, and we can
safely extrapolate that small variations from this distortion pattern 
can only correspond to variations in the parameters smaller than the differences shown in the table 
which were obtained comparing with using no correction for the non-radial distortion,
and most likely negligible.

\subsection{Position Angle P}

The roll angle of the SOHO spacecraft is maintained such that the effective
position angle\footnote{It is the position angle of the
northern extremity of the solar rotation axis with respect to MDI detector $y$-axis.}
($P_{\rm eff}$) of the MDI images should always be zero.
However, a 0.2$\Deg$ difference has been measured 
by intercomparing MDI and GONG images obtained in 1999 and 2000
(Cliff Toner, private communication).
\inlinecite{Beck05} estimated a 0.07$\Deg$ difference, assuming that there is no equator-crossing flow,
after re-gaining contact with the SOHO spacecraft (in 1999)
and 0.1$\Deg$ before losting contact.
No noticeable effect is seen in the observed ridge parameters after including a 0.2$\Deg$ correction
(Table~\ref{variation}).
We do not see the 3 nHz variation in the $a_1$-coefficient offset
that was predicted by our ridge model using a slightly higher correction of 0.25$\Deg$ 
(KRS).
This is probably because we did not accurately model the $P$-angle correction
in the leakage matrix calculation,
but added a very simple approximation of its effect.

\subsection{Carrington Elements}

Accordingly to \inlinecite{Beck05},
the standard values used for the two angles
specifying the orientation of the solar rotation axis $(i, \Omega)$\footnote{
$i$ is the angle between the plane of the ecliptic and the solar Equator
and $\Omega$ is the angle between the {\color{black}crossing} point of the solar Equator with the
ecliptic and the Vernal Equinox.}, known as the Carrington
elements, are off by $\Delta i = 0.095\Deg \pm 0.002\Deg$
and $\Delta\Omega = 0.17\Deg \pm 0.1\Deg$.
This introduces a time-varying correction in the calculation
of the rotation axis projection in the plane of the sky,
{\it i.e.}
the position angle, $P$, and the roll angle, $B_0$\footnote{$B_0$ is the heliographic latitude
of the central point of the disk and presents an annual variation.}. 
This correction in the $P$ angle
will be on top of the one mentioned in the previous section.  
Introducing a correction of $\Delta i = 0.1\Deg$ and $\Delta\Omega = 0.1\Deg$
in the image spatial decomposition
has no significant effect on the observed ridge parameters
(Table~\ref{variation}).

\section{The High-Degree Mode Parameters}
\label{highdegree}

Here we analyze the frequencies and splitting coefficients obtained using the high-$\ell$ data set.
The mode parameters were obtained after
including the five known instrumental effects in the spatial decomposition 
of the high-$\ell$ data sets described in Section~\ref{instreff},
fitting their observed power spectra using an asymmetric Lorentzian profile (Section~\ref{useddata})
and applying the ridge-to-mode correction to the fitted ridge parameters
(Section~\ref{ridgemodel}).

First, in Section~\ref{sub:check}, the high-$\ell$ data set mode frequencies
and splitting coefficients are compared with the values obtained by the medium-$\ell$ analysis.
Figure~\ref{fig:lnu} shows the region of common modes between the two sets. 
This comparison is done to check the goodness of the estimation of the mode parameters 
from the observed ridge in the high-$\ell$ data set
at these high medium-$\ell$ common modes.
Then, the estimated high-$\ell$ frequencies and splitting coefficients at $\ell \ge 100$ are analyzed
(Sections~\ref{hl_freq} and \ref{hl_ai}).

\subsection{Medium- and High-$\ell$ Set Comparison}
\label{sub:check}

\begin{figure}
\centerline{\includegraphics[width=0.8\textwidth,clip=]{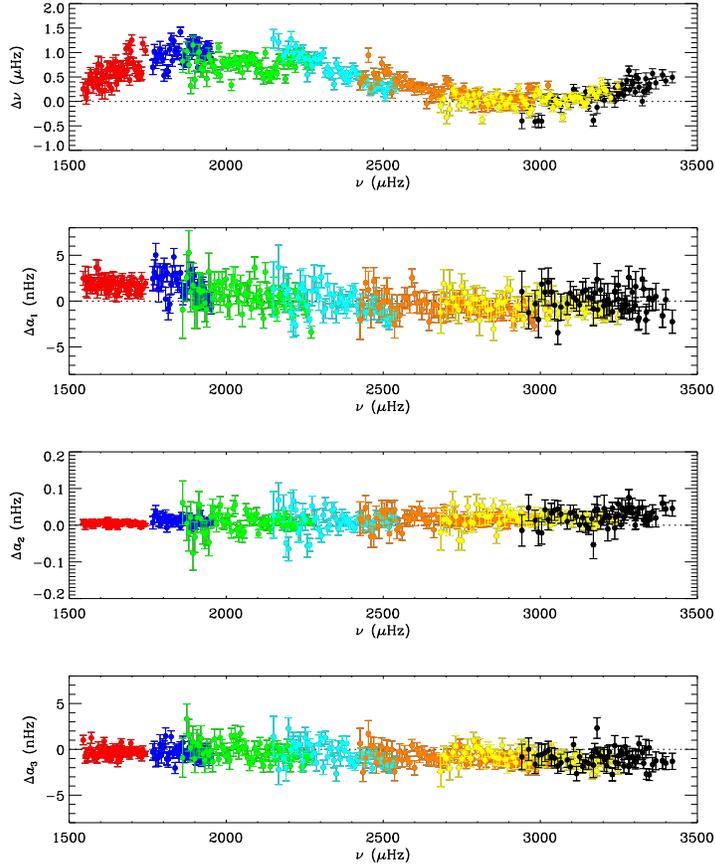}
	   }
\caption{
Central frequency and splitting coefficients difference between the 
medium- and high-$\ell$ sets observed during 2004 (Table~\ref{table_epochs}).
The differences are in the sense high- minus medium-$\ell$ set.
Modes with $n \le 5$ are in color where the $f$ modes are in red.
}
\label{fig:D04-top}
\end{figure}

Figure~\ref{fig:D04-top} shows the differences between the mode frequency and the splitting coefficients
obtained using the 72-day medium-$\ell$ data set described in Section~\ref{useddata} and 
using the high-$\ell$ set both observed during 2004 (Table~\ref{table_epochs}).
The high-$\ell$ set used in this comparison was analyzed as described in Sections~\ref{useddata},
\ref{ridgemodel}, and including all instrumental effects analyzed in the Section~\ref{instreff}.
In the medium-$\ell$ mode range,
the spatial leaks are well separated from the target mode and 
individual modes can be identified and fitted.
By decreasing the observed high-$\ell$ set frequency resolution using a short time
string (4096 minutes),
and thus increasing the width of the window function, 
we force the width of the spatial leaks to increase.
The spatial leaks in the high-$\ell$ power spectra now overlap with 
the target mode forming a ridge at $\ell$ as low as 100\footnote{An even
shorter time series, 2048 minutes, was used to check 
the results at the lowest high-$\ell$ modes.}
and the ridge-to-mode correction described in Section~\ref{ridgemodel}
is applied.
To increase the number of common modes,
the high-$\ell$ power spectra used in the comparison were fitted for every $\ell$ (and not every
tenth $\ell$).
The set of modes used in this comparison 
consists of 420 $p$ modes with $100 \le \ell < 200$
and $1.7 < \nu < 3.5$ mHz and 60 $f$-modes with $230 < \ell < 300$ and $1.5 < \nu < 1.8$ mHz
observed during 2004.

%.............................................................................
%.............................................................................

The mode frequencies agree within 1 $\mu$Hz
(Figure~\ref{fig:D04-top} top panel).
Their difference varies with frequency having a maximum at 2 mHz and decreasing to
zero (or near zero) at 1.6 and 3 mHz.
There is no indication of a degree dependence.
Although 1 \, $\mu$Hz is a small value, it is almost one order of magnitude {\color{black}larger} %higher 
than the fitting uncertainty of these modes (0.12 $\mu$Hz on average)
and hence undesirable. 
This seems to indicate that the frequency offsets $\Delta^\circ\nu$ {\color{black}(Figure~\ref{fig:offset})}
are underestimated 
by $\approx$35\% for modes with frequencies in the range 1.7 to 2.3 mHz,
this amount decreases to 10\% around 1.6 mHz and zero in the interval between 2.6 and 3.2 mHz where
the estimation is correct.
There is some indication that at 3.4 mHz the offsets are again underestimated (by $\sim$30\%).
Unfortunately, there are no common modes at higher frequencies.

A variation in the mode frequency that is purely a function of frequency
does not affect the outcome of the solar structure inversion, since it is
removed together with the near-surface errors in the physics of the solar model 
(see Section~\ref{hl_freq}).	
The instrumental effect that is probably causing the frequency offset to be underestimated 
at certain frequencies might be the instrumental PSF which is not included in the image spatial decomposition
or in the leakage matrix calculation (see Section~\ref{sec:imagescale}). 
Besides the frequency dependence, there is a puzzling frequency difference ($\approx 0.5 \mu$Hz)
in the $p_2$ modes 
(green circles in Figure~\ref{fig:D04-top} top panel)
with respect to the adjacent $n$ values.

The difference in the $a_1$ coefficients between the medium and high-$\ell$ set
is negligible except for $n = 0, 1$ modes (second panel in Figure~\ref{fig:D04-top}). Their mean difference
normalized by the fitting uncertainties 
is, in units of $\sigma$, $-0.2$ for $n \geq 2$, $2$ for $n =1$, and $3$ for the $f$ modes.
The horizontal-to-vertical displacement ratio decreases exponentially with $n$.
Thus, the differences observed only for $n = 0, 1$ modes 
could be an indication of an unaccounted instrumental effect that 
has a stronger effect on the horizontal component of the leakage matrix than on its vertical component.
Note also that the $f$ modes in the medium-$\ell$ set are fitted using
a different frequency interval around the peak in the power spectrum than the $p$ modes \cite{Schou99}.

There are small differences between the medium- and high-$\ell$ set 
for the splitting coefficients $a_2$ and $a_3$
(third and bottom panels in Figure~\ref{fig:D04-top}).
Their mean differences 
normalized by the fitting uncertainties of the high-$\ell$ power spectra
are
$a_2 = 1$ and $a_3 = -1$ in units of $\sigma$.

In conclusion, the estimation of the mode frequency and splitting coefficients 
from the observed ridge in the high-$\ell$ data set
at these moderate-degree values
is, in general, quite good. 
However, there is still room for improvement,
specially for the central frequency and the $f$-mode $a_1$ splitting coefficient.

The corrections in the medium-$\ell$ mode parameters calculated by \inlinecite{Larson08},
described in Section~\ref{useddata},
do not have a large influence in our results.
Their effect in Figure~\ref{fig:D04-top},
including the variation between fitting symmetric and asymmetric profiles,
is small in comparison
with the difference between the medium- and high-$\ell$ sets.
The mean total correction 
in the medium-$\ell$ frequency 
and splitting coefficients $a_1$, $a_2$ and $a_3$ to be applied to the values used here 
are: 0.130\,$\pm$\,0.004 $\mu$Hz, -0.244\,$\pm$\,0.009 nHz, 
(1.78\,$\pm$\,0.08)$\times10^{-3}$ nHz and -0.137\,$\pm$\,0.006 nHz respectively
\cite{Larson08}.
The average was calculated over the 420 
medium- and high-$\ell$ common modes.
The largest correction is in the central frequency.
The mean frequency variation between fitting symmetric and asymmetric profiles
is 0.043\,$\pm$\,0.002 $\mu$Hz for the common modes \cite{Larson08}.
The improved medium-$\ell$ analysis
will decrease slightly the differences between medium- and high-$\ell$ frequencies
showed in the top panel of Figure~\ref{fig:D04-top}.
However, it does not change the overall behavior of the frequency differences.

\subsection{The Central Frequency}
\label{hl_freq}

\begin{figure}
\centerline{\includegraphics[width=0.8\textwidth,clip=]{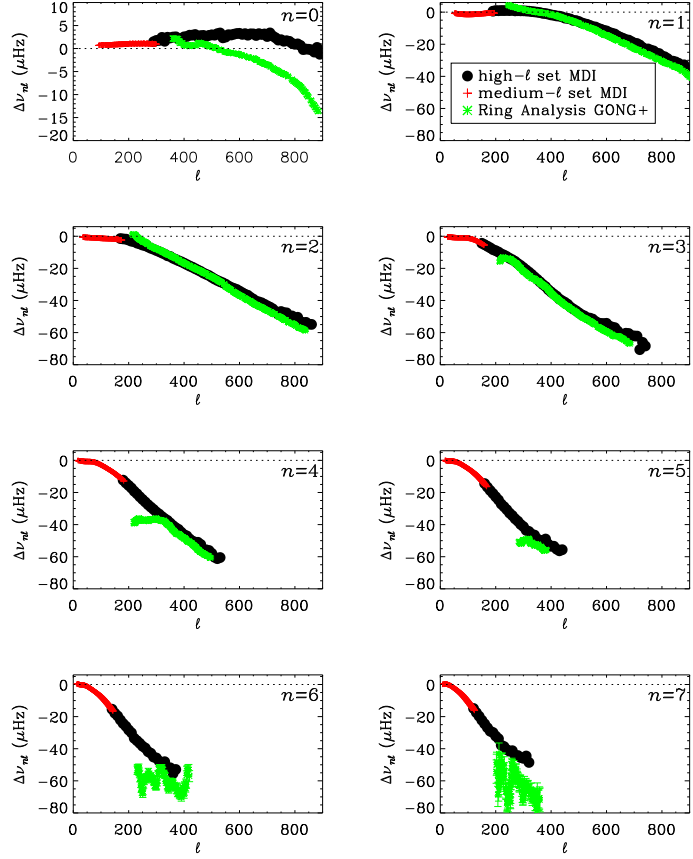}
	   }
\caption{
Difference between observed frequencies and their theoretical value
as a function of degree for different $n$ values using the high-$\ell$ set data obtained in 2004 (in black).
The differences are in the sense observed minus theoretical.
For comparison, the results for the medium-$\ell$ set (red crosses) and from ring-diagram analysis (green stars) are also plotted.
The fitting uncertainties are given by the error bars.
Note the different scale for the $f$ modes.
}
\label{fig:rings}
\end{figure}

Figure~\ref{fig:rings} shows the difference between the high-$\ell$ set mode frequencies 
($100 \le \ell \le 900$) and their theoretical value calculated from Christensen-Dalsgaard's
model S \cite{CD96} as a function of degree for different $n$ values.
The high-$\ell$ set used in this comparison was analyzed as described in Sections~\ref{useddata},
\ref{ridgemodel}, and including all instrumental effects analyzed in the Section~\ref{instreff}.
The mode range is shown in Figure~\ref{fig:lnu}.
The medium-$\ell$ set frequencies are also plotted as a reference.
In the absence of any acceptable theory to describe the physics of the layers near the
solar photosphere, the difference between the observed and theoretical frequencies
due to the near-surface errors in the model are well known to be quite large
({\it e.g.}, \opencite{CDCox}).
The general trend is such that the observed $p$-mode frequencies are smaller than their theoretical prediction and,
at a high enough degree, the differences increase with degree and with frequency. 
Accordingly to the results obtained by the high-$\ell$ set, this difference can be as large as 60 $\mu$Hz.
In fact, for a given $n$, the frequency differences increase almost linearly with degree 
with a slope that also increases linearly with $n$.
The high-$\ell$ set analysis
gives consistent frequencies for all observational periods listed in Table~\ref{table_epochs},
where the differences in the frequencies obtained at the different observational periods
can be explained by their well known variation with solar cycle activity
and it is described in detail in Section~\ref{solarcycle}.

In order to compensate for the frequency shifts due to the near-surface errors in the model, an
unknown function ($F_{{\tiny\mbox{surf}}}$), the so-called surface term ($F_{{\tiny\mbox{surf}}}$) 
is usually added to the equation governing helioseismic inversions: 
$F_{{\tiny\mbox{surf}}} = Q_{n,\ell} \times (\delta\nu_{n,\ell}/\nu_{n,\ell})_{{\tiny\mbox{surf}}}$,
where $Q_{n,\ell}$ is the mode inertia normalized by the inertia of a radial mode of the same frequency.
\inlinecite{CD89} from the asymptotic theory of solar $p$ modes pointed out that
low- and moderate-degree modes propagate nearly vertically near the surface; thus, their behavior
in this region is essentially independent of degree and depends only on frequency.
This however does not hold for high-degree modes as can be inferred from Figure~\ref{fig:rings}.
A second-order asymptotic approximation,
where the surface term is a function not only of frequency but also of $L/\nu_{n,\ell}$
\cite{brodsky93}, must be used as shown by \inlinecite{DiMauro2002}.

%.............................................................................
%.............................................................................

Figure~\ref{fig:rings} also shows for comparison the frequency difference 
applying the ring-analysis technique obtained by \inlinecite{RBB07}.
A magnetically quiet region (15$\Deg$ in diameter) observed by the GONG+ network 
on 4 July 2005 during Carrington rotation 2031 centered at 
115$\circ$ in longitude and 3$\circ$ south in latitude
was tracked for 8192 minutes 
at the appropriate photospheric rotation rate.
The region crossed the central meridian at the middle of its tracking interval.
The tracked region was mapped to a plane using Postel's projection and 
its power spectra, given by the three-dimensional Fourier transform of the temporal series of images,
were fitted using the 13-parameter model of \inlinecite{Basu99}.
The wavenumber ($k$) can be identified with the degree of a spherical harmonic mode of global
oscillations by $L = k R_{\odot}$.
As the oscillations in a plane-parallel geometry are only discrete in radial order,
$\ell$ does not need to be an integer. 
Each ``mode'' is obtained by fitting a region of power spectrum that has significant overlap with
those covered by neighboring ``modes'', making them not strictly independent.
To confirm that the frequencies obtained for this particular solar region
correspond to typical values, they were compared with the frequencies obtained for 
13 additional quiet regions, with latitudes ranging from -12$\Deg$ to +12$\Deg$ \cite{RBB07}.
Their difference is smaller than 
3 $\mu$Hz for $n \leq 5$. 
Part of this variation is due to the fact that
the projection of the spherical solar surface onto a flat detector introduces some 
foreshortening that depends on the distance of the region from disk center, which
can introduce systematic errors in determining the mode characteristics.

The frequency differences (in relation to the theoretical value) obtained using ring analysis
present the same trend as using spherical harmonic decomposition,
except for the $f$ modes.
In most cases, the ring-analysis frequencies are smaller than the ones obtained by the high-$\ell$ set.
For $p$ modes, their difference 
in the sense global minus ring analysis
varies between -4 and 6 $\mu$Hz for $n \leq 5$
and it could be as large as 20 $\mu$Hz for $n > 5$.
Due to the small size of the region analyzed, the ring-analysis power spectra
have low spatial resolution doing poorly at medium degree and 
at high $n$, as can be seen for $n = 6, 7$ in Figure~\ref{fig:rings}.
Note also that for $n = 4$ and $\ell \le 300$ modes in Figure~\ref{fig:rings}, the difference between the ring-analysis 
and theoretical frequencies is constant and does not continue to decrease as $\ell$ decreases.
This is probably an indication of a {\color{black}poor} %bad 
frequency determination by the ring analysis at these medium-$\ell$ values.
{\color{black}In} %As a 
conclusion, for the $p$ modes, the ring-analysis frequencies agree with the global-analysis values
within $\pm 6$ $\mu$Hz, 
which is much smaller than the difference between either of them and the frequencies 
obtained from the solar model.

For the $f$ modes, 
the frequency differences (in relation to the theoretical value) obtained using ring and global analysis
are different.
The medium-$\ell$ $f$-mode frequency is on average $0.9 \pm 0.1$ $\mu$Hz larger than the theoretical values.
The $f$-mode frequency obtained by ring analysis for $\ell \le 450$
is on average $1.1 \pm 0.6$ 
$\mu$Hz larger than the theoretical frequency. 
For $\ell > 450$, it decreases sharply with degree 
and it
can be described by
a {\color{black}fifth} degree polynomial\footnote{The polynomial in $\ell$ has the following coefficients
in units of $\mu$Hz:
$c_0 = 350\pm80$, $c_1 = -2.7\pm0.7$, $c_2 = 0.008\pm0.002$, $c_3 = (-1.3\pm0.3)\times10^{-5}$, 
$c_4 = (1\pm0.2)\times10^{-8}$ and $c_5 = (-3.2\pm0.7)\times10^{-12}$.}.
At $\ell = 880$, it is 13 $\mu$Hz smaller than the theoretical values.
Such a sharp decrease in the $f$-mode frequency in relation to the theoretical values similar
to the one observed for the $p$-modes 
was already reported by other authors (\opencite{Duvall98} and references within).
The frequencies of the $f$ modes, contrary to the $p$ modes, depend only weakly on the hydrostatic structure of the model
({\it e.g.} \opencite{Gough93})
and a possible explanation is that the $f$-mode frequencies are reduced by granulation \cite{Murawski93}.
However, 
the $f$-mode frequency obtained by global analysis presents a different trend in relation 
to the theoretical frequency.
For $\ell < 710$, it is only
2.7\,$\pm$\,0.5 $\mu$Hz on average larger than the theoretical frequency. 
At larger $\ell$, it decreases linearly with degree with a slope of $-0.021 \pm 0.001$ $\mu$Hz;
it is zero at $\ell = 850$.
The global-analysis results suggest
a much weaker interaction between the $f$ modes and the granulation
than predicted by the ring analysis.

The error bars in Figure 6, given by the fitting uncertainties
are too small to be seen, except for the ring-analysis frequencies at $n \ge 6$.
The observed frequency errors
obtained using the medium-$\ell$ and high-$\ell$ sets and ring analysis
are in the range 0.007\,--\,0.4 $\mu$Hz, 0.06\,--\,0.6 $\mu$Hz and 0.2\,--\,15 $\mu$Hz respectively.
The ratio between the global- and ring-analysis fitting uncertainties varies from
0.9 to 33.
The ring-analysis uncertainties are similar to those of global analysis at low $n$
($n \le 2$) and high $\ell$ ($\ell > 700$). 
They become much higher than the global-analysis uncertainties
as $n$ increases and as $\ell$ decreases.
The frequency is not a parameter in the model used to fit the 
ring-analysis power spectra (see \opencite{Basu99}), 
the observational error estimated for the fitted width 
is used instead, following \inlinecite{Rajaguru01}.

\subsection{The Rotational Splitting Coefficients}
\label{hl_ai}

High-$\ell$ splittings can be used to infer the solar rotation rate in the outermost layers
of the solar convection zone.  
At these layers, there is a steep gradient of the rotation,
making it a very interesting region to study.
This was first suggested by the fact that different surface markers
such as sunspots, faculae, H$\alpha$ filaments, and supergranular network
present different rotation rates ({\it e.g.} \opencite{Snodgrass92}) which has been interpreted
by assuming that the markers are anchored
at different depths \cite{Foukal}.
The solar rotation rate determined by helioseismology (using modes with $\ell \le 300$)
has a local maximum at about 0.95$R_\odot$
(35 Mm below the solar surface) and decreases fast towards the surface
\cite{Howe07}.
Using $f$-mode splittings with degree between 117 and 300, 
\inlinecite{Corbard02} estimated a constant radial gradient of solar angular velocity of around
$-400$ nHz/$R_\odot$ in the outer 15 Mm between the Equator and 30$\Deg$ latitude. Above 30$\Deg$
it decreases in absolute magnitude to zero around 50$\Deg$.
The addition of high-degree splittings to this analysis will help to constrain the
determination of the solar rotation profile
both closer to the solar surface and at higher latitudes.

\begin{figure}
\centerline{\includegraphics[width=0.8\textwidth,clip=]{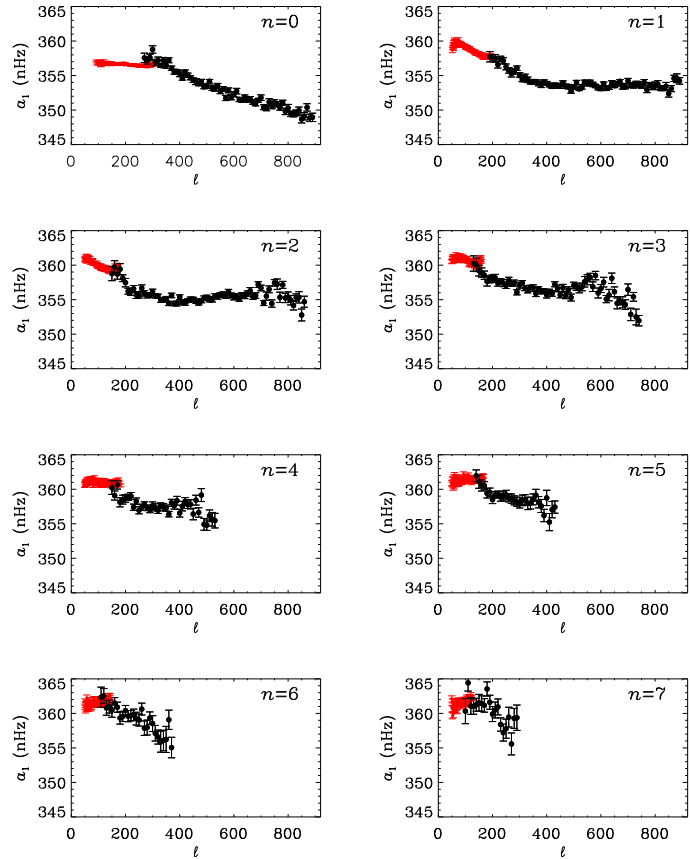}
	   }
\caption{The $a_1$
rotational-splitting coefficients as a function of degree for the 2004 data set,
where the ones obtained with the medium-$\ell$ set are in red and 
with the high-$\ell$ set in black.
The fitting uncertainties are given by the error bars.
}
\label{fig:D04-a1}
\end{figure}

Figure~\ref{fig:D04-a1} shows a smooth variation with degree of the high-$\ell$ 
set $a_1$ coefficients as expected, implying a good estimation of the mode splittings.
For a given $n$, the $p$-mode $a_1$ coefficients decrease with degree until 
$\ell \approx 300$ by 1\,--\,2\% (which corresponds to 4\,--\,6 nHz) 
where it becomes approximately constant. 
And the $f$-mode $a_1$-coefficients decrease almost linearly
with degree from $\ell \approx 300$ to 900 by 2.5\% (or 9 nHz).
For a given $n$, the $a_1$ variation with $\ell$ shown in Figure~\ref{fig:D04-a1}
presents the same trend if plotted against the 
location of the mode lower turning point (given by $L/\nu$),
indicating the expected decrease of the solar rotation with solar radius near the solar surface.
The $p$-mode $a_1$-coefficients, {\color{black}especially} for $2 \le n \le 5$, present
a sharp decrease near $\ell = 200$ correspoding to a turning point of 
depth of $\approx$30 Mm, which is at the beginning of the solar rotation {\color{black}decrease}. %drop.

Although the medium-$\ell$ $a_3$-coefficient decreases slightly in absolute value with
degree (from -7.8 nHz at $\ell\approx50$ to -7.3 nHz at $\ell\approx300$),
the high-$\ell$ $a_3$-coefficient mean value 
(-8.6\,$\pm$\,1.2 nHz)
is larger in absolute value than the mean medium-$\ell$ value (-7.5\,$\pm$\,0.3 nHz).
The high-$\ell$ $a_3$-coefficient does not show any variation with degree.
However, it has a small absolute value at low frequency ($\nu \le 2.2$ mHz), 
which is similar to the medium-$\ell$ mean: -7.4$\pm$0.5 nHz.

While the odd coefficients provide information about the internal solar rotation rate, 
the even coefficients arise from latitudinal structural variation, centrifugal distortion, and
magnetic fields.
The $a_2$-coefficients obtained with the high-$\ell$ set are on average zero
($1500 < \nu < 5200$ $\mu$Hz). 
The $a_2$-coefficients obtained with the medium-$\ell$ set are on average zero
for $\nu < 2$ mHz. At higher frequencies, 
they start to decrease with frequency,
where for $\nu=3550\pm50$ $\mu$Hz, its mean value 
is -0.08\,$\pm$\,0.03 nHz.

%.............................................................................
%.............................................................................

\section{Frequency Variation Over the Solar Cycle}
\label{solarcycle}

The correlation between solar acoustic mode frequencies and the magnetic
activity cycle is well established and has been substantially studied during
the last and current solar cycles (see for example \opencite{Cha07}, \opencite{DG05} and references therein).  
However, its physical origin is still a matter of debate
and the detailed analysis of the frequency shift characteristics
are likely to contribute {\color{black}to} %in 
solving this problem.
The data sets used in this work and listed in Table~\ref{table_epochs}
cover a considerable part of solar cycle 23.
They also cover a large degree range ($20 \leq \ell \leq 900$) when combining the medium- and high-$\ell$ sets.
In a previous work \cite{RKS08}, we used these two data sets to analyze the characteristics of
the frequency variation along the solar cycle.
Here we improved our previous analysis
by calculating the probability that a linear relationship indeed exists
between the frequency shift of a given $(n,\ell)$ mode and the solar-activity index.
Although the medium- and high-$\ell$ sets are fitted with symmetric and asymmetric profiles respectively,
the resulting difference in frequency does not change 
on average
over the solar cycle \cite{Larson08}.

The frequency variation with the solar cycle is
given by the frequency difference between 1999\,--\,2004 epochs with respect to the 2005 epoch,
the one with the lowest activity index in our data set.
For each $(n,\ell)$ mode, their frequency shifts were fitted 
assuming a linear relationship with the solar activity index with a zero intercept
and using a weighted least-squares minimization.
For a detailed description, see \inlinecite{RKS08}.
Four different solar activity indices 
commonly used in the literature and that are available in the period 1999\,--\,2005
were used in this analysis: 
the solar UV spectral irradiance\footnote{\url{http://www.ngdc.noaa.gov/stp/SOLAR/ftpsolaruv.html}}
(given by the NOAA Mg {\sc ii} core-to-wing ratio),
the Magnetic Plage Strength Index (MPSI\footnote{\url{http://www.astro.ucla.edu/~obs/intro.html}},
Mt.\ Wilson Observatory),
the solar-radio 10.7-cm flux\footnote{\url{http://www.ngdc.noaa.gov/stp/SOLAR/ftpsolarradio.html}}
(National Research Council of Canada) and
the sunspot number (SSN\footnote{\url{http://sidc.oma.be}}, SIDC, RWC, Belgium).
The minimum-to-maximum solar cycle frequency shift ($\delta\nu^e_{n,\ell}$) was estimated by
multiplying the slope of the linear fit by the corresponding solar index 
variation between the maximum and minimum of cycle 23.
The mean activity level for each epoch 
in relation to the maximum reached in January 2002 
is listed in Table~\ref{table_epochs},
given by the solar UV  spectral irradiance.

In an attempt to include in the analysis of the solar-cycle induced frequency shifts
only modes that are, in fact, correlated with the solar activity,
\inlinecite{RKS08} rejected modes whose Pearson correlation coefficient
is smaller than 0.8, or whose slope uncertainty is larger than 20\% of its absolute value.
However, the correlation coefficient cannot be used directly to indicate 
the probability that a linear relationship exists between two observed quantities,
{\it i.e.}, if, indeed, there exists a physical linear relation between them.
A small correlation coefficient might indicate only a small slope in their 
linear relation.
A better approach is to calculate the probability that the data points 
represent a sample derived from an uncorrelated parent population.
We estimated the probability $P_u(r)$ that a random sample of $N$ uncorrelated data points would yield
a linear-correlation coefficient as large as or larger than the observed absolute
value of $r$ \cite{Bevington92}.
If $P_u(r)$ is very small, it would indicate that it is very improbable that they are linearly
uncorrelated.
Thus, the probability is high that the frequency shift of a particular $(n,\ell)$ mode
and the solar-activity index used are correlated and the linear fit is justified.

\begin{figure}
\centerline{\includegraphics[width=0.8\textwidth,clip=]{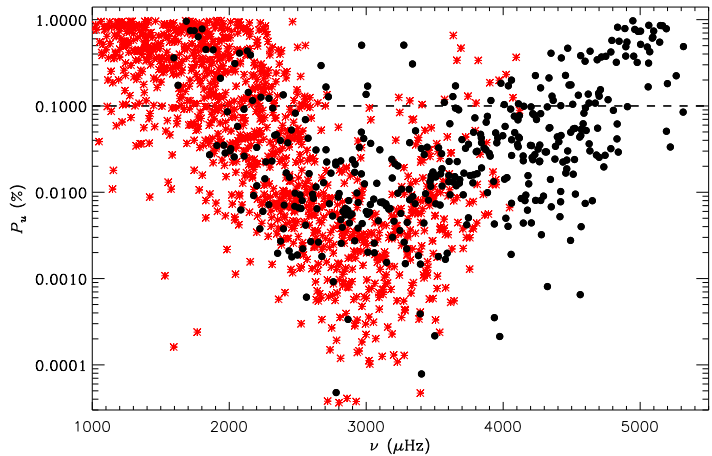}
	   }
   \caption{
Probability $P_u(n,\ell;i)$
as a function of the mode frequency for the medium (red) and high-$\ell$ (black) sets,
using the sunspot number as the solar activity index $i$.
}
\label{fig:pc}
\end{figure}

For each $(n,\ell)$ mode and solar-activity index $i$, we calculated the Pearson 
correlation coefficient $r(n,\ell;i)$ and the probability $P_u(r) \equiv P_u(n,\ell;i)$.
Figure~\ref{fig:pc} shows the probability $P_u(n,\ell;i)$ as function of the mode frequency.
It is very similar for all four solar-activity indices.
Modes with frequency around 3 mHz have the highest probability of being correlated.
It decreases for modes with frequency smaller than $\approx$2.5 mHz or larger than $\approx$4.5 mHz.
There is no noticeable dependence of $P_u$ with $\ell$ or $\nu/\ell$.

%---------------------------------------------------------
%---------------------------------------------------------

\inlinecite{LW90} were the first ones to suggest
that the solar-cycle frequency variation ($\delta\nu_{n,\ell}$)
is linearly proportional to the inverse mode inertia ($I_{n,\ell}$)
using observations obtained at the Big Bear Solar Observatory.
The observed frequency shift is larger for higher-frequency and for higher-$\ell$ modes.
As pointed out by \inlinecite{LW90}, these modes are more sensitive to surface perturbations, 
because they have, respectively, higher upper- and lower-reflection points in the Sun.
This
indicates that the dominant structural changes during
the solar cycle, inasmuch as they affect $p$-mode frequencies, occur near the solar surface.
\inlinecite{Cha01} 
obtained
a similar shift at low frequencies ($\nu \le 2.5$ mHz): $\delta\nu_{n,\ell} \propto 1 / I_{n,\ell}$.
However, at higher frequencies, they found: $\delta\nu_{n,\ell} \propto \nu^{\alpha} / I_{n,\ell}$,
where $\alpha = 1.91 \pm 0.03$.
They used data sets with a much higher duty cycle obtained
by the ground-based networks, GONG and BiSON, and the SOHO satellite (VIRGO/LOI). 

The value of $\alpha$ depends upon the physical mechanism responsible for affecting the acoustic mode frequencies
during the solar cycle.
Several authors (\opencite{Gough90}; \opencite{LW90}; \opencite{Goldreich91}) 
estimated $\alpha = 3$ assuming modifications in the thermal structure of the Sun 
in a thin layer at the photosphere.
On the other hand, \inlinecite{Gough90} pointed out that if 
the frequency shifts are caused by
variations in the efficacy of the convection during the solar cycle,
$\alpha \approx -1$.

\inlinecite{RKS08} showed that the scaled frequency shift can be described with a simple power law
at all frequencies:
\begin{equation}
 \delta\nu^e_{n,\ell} = C_{\gamma} ~
\frac{(\nu_{n,\ell})^{\gamma}}{Q_{n,\ell}},
   \label{eq:plaw}
\end{equation}
where $Q_{n,\ell}$ is the mode inertia, normalized by the inertia of a radial mode of the same frequency,
$I_{\ell=0}(\nu_{n,\ell})$,
calculated from Christensen-Dalsgaard's model S (see \opencite{CD96}).
The $f$ and $p$ modes were fitted independently of one another.
\inlinecite{Cha01} also plotted the frequency shift scaled by the normalized mode inertia (upper right-hand panel
of their Figure~1). However, they {\color{black}chose} to fit the frequency shift scaled by the mode inertia instead.
The multiplication of the frequency shift by $Q_{n,\ell}$ normalizes 
the shift to its expected radial equivalent.

\begin{figure}
\centerline{\includegraphics[width=0.8\textwidth,clip=]{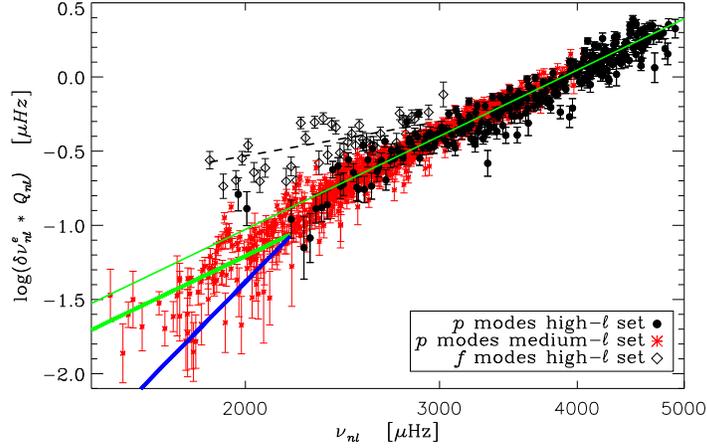}
	   }
   \caption{Frequency shift multiplied by the normalized
mode inertia as a function of frequency obtained, in this case, using 
the sunspot number as a proxy of solar activity.
The long green continuous line is the fit to the all $p$ modes with $\nu \geq 2.5$ mHz and
the dashed black line is to the $f$ modes in the high-$\ell$ set.
The thick green short line {\color{black}is the fit to $p$ modes with $\nu \leq 2.3$ mHz using 
the same slope as the long green line but a different $y$-intercept.}
The blue line corresponds to a frequency shift given by $\alpha = 0$
with an arbitrarily chosen $y$-intercept.
}
\label{fig:slope}
\end{figure}

\begin{figure}
\centerline{\includegraphics[width=0.8\textwidth,clip=]{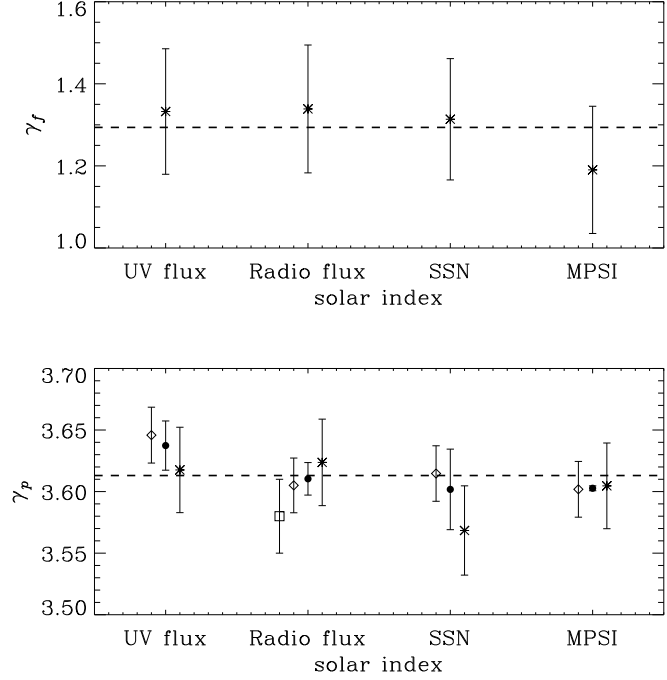}
	   }
   \caption{Exponent $\gamma$ obtained fitting Equation~(\ref{eq:plaw}) to the $f$ (top)
and $p$ (bottom) modes for four different solar indices. 
In the top panel, the exponents were obtained using only the high-$\ell$ set.
In the bottom panel, the diamonds and stars were obtained using the medium- and high-$\ell$ sets respectively 
and their {\color{black}weighted} average is given by the circles. 
The square is the $\gamma$ corresponding to the $\alpha$ value quoted in 
Chaplin {\it et al.} (2001)
for the high-frequency interval.
The fitting uncertainties are given by the error bars.
The dashed lines are the {\color{black}weighted} average between the four solar indices.
}
\label{fig:gamma}
\end{figure}

As in \inlinecite{RKS08}, Equation~(\ref{eq:plaw}) was fitted to estimate $\gamma_p$ (for the $p$ modes)
and $\gamma_f$ ($f$ modes),
using a weighted least-squares minimization 
(thin green line and black dashed line in Figure~\ref{fig:slope} respectively).
Only modes that have a probability of 0.1\% or less that its variation is
linearly uncorrelated with the solar index were included in the present analysis.
They represent 80\% of the high-$\ell$ set modes and 65\% of the medium-$\ell$ set modes.
The frequency ranges used in the fitting of the medium- and high-$\ell$ $p$ modes and high-$\ell$ $f$ modes
are: 2.5\,--\,4.1 mHz, 2.8\,--\,4.9 mHz and 1.8\,--\,3.0 mHz respectively.
It seems that there is a step in the $p$-mode frequency shift around 2.3 mHz and only modes with $\nu \ge 2.5$ mHz were included
in the fitting of the medium-$\ell$ set.
$p$-modes with $\nu \le 2.3$ mHz
seem to have a similar slope to the fitted high-frequency modes,
but a different $y$-intercept.
This is illustrated by the thick green short line in Figure~\ref{fig:slope}.
The fact that the $f$ modes are affected by the solar cycle in a different way
than the $p$ modes is expected since they have very different properties.
{\color{black} The $f$-mode is essentially a surface wave and 
its frequency is less likely to be influenced by the solar stratification than the $p$-mode frequency.}
We repeated the analysis for all four solar indices.
The results are plotted in {\color{black}Figure~\ref{fig:gamma}}.
The exponents obtained using the different solar indices agree within their
fitting uncertainty. 
The {\color{black}weighted} mean from the four solar indices 
is $\bar{\gamma_f} = 1.29 \pm 0.07$ for the $f$ modes and
{\color{black}$\bar{\gamma_p} = 3.60 \pm 0.01$} for the $p$ modes.  %$\bar{\gamma_p} = 3.61 \pm 0.02$ for the $p$ modes.
$\bar{\gamma_p}$ was calculated after averaging 
the values obtained for the medium- and high-$\ell$ sets (given by the circles in Figure~\ref{fig:gamma}).
Only $f$ modes obtained with the high-$\ell$ set were used to estimate $\gamma_f$.
There are only a few modes (seven) in the medium-$\ell$ set with a probability
of 0.1\% or less of being uncorrelated, they have a very small frequency shift 
{\color{black}($0.05 \pm 0.01$ $\mu$Hz on average)}, 
and by consequence they were excluded from the fit. 
The $f$ modes in the medium-$\ell$ set have very small frequencies ($\nu < 1.5$ mHz)
and, accordingly to Figure~\ref{fig:pc}, low-frequency modes are not well correlated with the solar cycle.
{\color{black}
High-$\ell$ $f$-modes with $\nu \leq 1.8$ mHz are also not well correlated with the solar cycle
and were excluded from the analysis. Their frequency shifts show a steep increase 
from values similar to those of the well-correlated medium-$\ell$ $f$-modes at 1.7 mHz
to values similar to the diamonds in Figure~\ref{fig:slope} at 1.8 mHz.
}

%---------------------------------------------------------
%---------------------------------------------------------

In \inlinecite{RKS08} only the solar UV spectral irradiance was used as a proxy for
the solar-cycle index to calculate the exponent $\gamma$.
The value obtained here for the $p$ modes,
using the same solar index but a better criteria for selecting the modes included in
the analysis, is the same as before: {\color{black}$\gamma_p = 3.63 \pm 0.02$}. However,
the value obtained for the $f$ modes ($\gamma_f = 1.33 \pm 0.2$) is 20\% smaller
than our previous result.
The distinct mode selection criteria account for this difference.
{\color{black}
As the number of $f$ modes is small (45), the fitting of
$\gamma_f$ is more sensitive to the mode selection.
}

To compare the above mentioned values of $\alpha$ obtained by \inlinecite{Cha01} with our results,
$\nu^{\alpha}_{n,\ell}$ was divided by $I_{\ell=0}(\nu_{n,\ell})$ and then fitted by Equation~(\ref{eq:plaw}) 
to estimate the corresponding $\gamma$ exponent.
The $\alpha$ values obtained at the two frequency intervals 
$1.6 < \nu < 2.5$ mHz ($\alpha = 0$)
and $2.5 < \nu < 3.9$ mHz ($\alpha = 1.91$)
correspond to $\gamma = 7.59 \pm 0.18$ and $\gamma = 3.58 \pm 0.03$ respectively.
Accordingly to the authors, the division of the data into two parts has been made on
a purely ``artificial'' basis. \inlinecite{RKS08} showed that 
this imposed ``breakpoint'' can be explained by the fact that
$\log(\delta\nu^e_{n,\ell} ~ I_{n,\ell})$ has a quadratic dependence on
$\log(\nu_{n,\ell})$, with an inflection point at 2.59 mHz
(see Figure~6 in \opencite{RKS08}).
\inlinecite{Cha01} used 10.7-cm radio flux as the solar activity index.
Their results at the high frequency interval agrees well with ours (represented by a square in Figure~\ref{fig:gamma} bottom panel).
However, at low frequency, their result is twice as large as the one obtained here
(blue line in Figure~\ref{fig:slope}).
The difference could be due to the mode selection used to obtain our results.
\inlinecite{DG05} used also a simple frequency difference including all modes.
They fitted $\delta\nu^e_{n,\ell} ~ I_{n,\ell}$ using truncated Legendre
polynomial series and 
fitted the $f$ and $p$ modes independently of one another.
For $p$ modes, their fitting agrees with \inlinecite{Cha01} thus disagreeing with ours
at low frequency.
For $f$ modes, in the frequency range 
{\color{black}
1.37\,--\,1.60 mHz,
the corresponding $\gamma_f$ (inferring from their Fig.~2) is 
five times large as our determination at $\nu > 1.8$ mHz,
and it is similar to the low-frequency $\gamma_p$ obtained by \inlinecite{Cha01}.}
\inlinecite{DG05} noted that 
the frequency shifts normalized by $I_{n,\ell}$ 
present an opposite trend for the $f$ and $p$ modes.
They increase with increasing frequency for $p$-modes and decrease for $f$ modes.
This opposite trend disappears when the frequency shifts are normalized by $Q_{n,\ell}$
instead.

%---------------------------------------------------------
%---------------------------------------------------------

Several authors have reported a sharp decrease of the frequency shifts at high frequency.
The positive shifts suddenly drop to zero, and become negative reaching absolute values
much larger than the positive shifts at moderate frequency. 
The falloff was reported to happen around 3.7 mHz by 
\inlinecite{Jefferies98} -- using $100 \leq \ell \leq 250$ modes - and
\inlinecite{Salabert04} -- using $\ell \leq 3$ modes,
which is supported by \inlinecite{LW90} determinations ($5 \leq \ell \leq 60$).
However, \inlinecite{Howe02} 
do not see any falloff for $\nu \leq 4$ mHz modes ($\ell \leq 300$) and
\inlinecite{Rhodes02} see the drop at 5 mHz (using $\ell \leq 1000$ modes).
The exact frequencies where
the frequency shifts become zero or negative or reach a maximum negative value also varies widely
between the publications.
This sharp decrease in the frequency shift has been associated to an increase in the
chromospheric temperature ({\it e.g.}, \opencite{Goldreich91} and \opencite{Jain96}).
From Figure~\ref{fig:pc}, high-frequency modes have a large probability $P_u$
of being uncorrelated with the solar cycle, at least until 5.2 mHz. 
Note that $P_u(r)$ is calculated for the absolute
value of $r$, estimating the probability of being correlated or anti-correlated with solar cycle.
In our analysis, the frequency shift drops sharply around 4.6 mHz and there are a few
modes ($\approx$ ten) with a negative frequency shift with frequencies between 5 and 5.35 mHz.
However, only modes with frequency smaller than 4.9 mHz obey the mode selection criteria
({\it i.e.}, $P_u \leq 0.1\%$) and were included in Figure~\ref{fig:slope}. The frequency shift
drop from 4.6 to 4.9 mHz is not seen in Figure~\ref{fig:slope} because of the mode-inertia normalization.

As first suggested by \inlinecite{LW90}, the most significant sources of frequency shift must be
localized near the solar surface since the frequency shifts seem to be independent of $\ell$.
Consider a small perturbation in the solar equilibrium model
localized near the solar surface
and the corresponding
changes in the mode frequency, $\delta\nu_{n,\ell}$.
For modes extending substantially more deeply than the region of the perturbation, the eigenfunctions
are nearly independent of $\ell$ at fixed frequency in that region 
(see Figure~8 and the associated discussion in \opencite{CDCox}).
From this, it can be inferred that if the quantity
$\delta\nu_{n,\ell} Q_{n,\ell}$ is independent of $\ell$ at fixed $\nu$ for a given set of modes,
then the perturbation is probably largely localized outside the radius given by the maximum
lower turning point of the set of modes considered \cite{CDCox}.
In the case of the frequency shift induced by the solar cycle, $\delta\nu_{n,\ell} Q_{n,\ell}$ is 
independent of $\ell$ at fixed $\nu$ for all modes analyzed here, where $\ell \le 900$ and $\nu/L > 4.2$ $\mu$Hz.
Thus, the perturbation causing the frequency shift is probably localized in the region
4 Mm below the solar surface or less.
Using the same reasoning,
the thickness of the near-surface region where
the uncertainties in the physics of the model 
are confined can be inferred.
In this case, the set of modes where
$\delta\nu_{n,\ell} Q_{n,\ell}$ is independent of $\ell$ at fixed $\nu$ 
is such that $\nu/L > 12$ $\mu$Hz and $\ell \le 370$ using the high-$\ell$ 
frequency determination (Section~\ref{hl_freq}).
Including higher-degree modes, $\delta\nu_{n,\ell} Q_{n,\ell}$ is not a function of
frequency alone anymore (see, for example, Figure~3 in \opencite{Rabello-Soares00}).
As a result,
the frequency uncertainties in the model are probably largely restricted
to the layers 
18 Mm below the surface.

\section{Conclusion} 
      \label{S-Conclusion} 

In the determination of unbiased high-degree mode parameters,
the instrumental characteristics must be taken into account in the image spatial decomposition
or in the leakage matrix calculation itself to obtain a correct estimation of the relative
amplitude of the spatial leaks.
Among the instrumental characteristics analyzed here, the image scale is the one that affects
the parameter determination the most. 
The image scale is the ratio of the image dimensions observed on the CCD detector and the dimensions in the actual Sun.
An error in the image scale introduces an error in the estimated central frequency which
increases with the mode frequency.
A 0.27\% error in the image scale would shift the estimated central frequency by
as much as 11 $\mu$Hz at 5 mHz. The radial distortion also has an important effect which is expected since it is 
similar to an image scale error. 
An instrumental property not taken into account here
(due to a lack of a good estimation)
that could have an important effect on the measured parameters is an azimuthally varying PSF.

The applied ridge-to-mode correction recovers frequencies at moderate degree that differ
from the assumed corrected values by 1 $\mu$Hz or less depending on the mode frequency.
The fitting uncertainty of the recovered frequencies is in the range 0.07\,--\,0.18 $\mu$Hz.
The agreement for the $a_1$, $a_2$, and $a_3$ coefficients is very good, except maybe for 
$f$- and $p_1$-mode $a_1$ coefficients, their
mean difference with respect to the assumed correct values is,
respectively, three and two normalized by the ridge fitting uncertainties. 

At high degree, the differences between our frequency determination and theoretical frequencies
for the $p$ modes
{\color{black}shows} the same general variation with degree as the results obtained with ring analysis.
For $n \leq 5$, the global and ring analysis agree within 6 $\mu$Hz.
The high-degree $f$-mode frequencies obtained using ring analysis, like previous observations
(\opencite{Duvall98} and references within), are substantially lower than the theoretical frequencies.
Surprisingly, the $f$-mode high-$\ell$ set frequencies agree well with the model frequencies (within 3 $\mu$Hz)
whereas the ring-analysis frequency differences can be as large as 13 $\mu$Hz for $\ell > 700$ modes.
The implications of the high-$\ell$ frequencies and splitting coefficients on the
solar structure and rotation will be addressed in a future paper.

{\color{black}As noted} %As noticed 
by other authors for low- and moderate-degree modes ({\it e.g.}, \opencite{LW90}), 
the frequency shift induced by the solar cycle scales well with the mode inertia.
We extended this analysis to high-degree modes and found out that scaling with
the mode inertia normalized by the
inertia of a radial mode of the same frequency
follows a simple power law
(given by Equation~\ref{eq:plaw}) with one exponent at all frequency ranges,
where the $f$ and $p$ modes are fitted independently of one another.
The exponents obtained using four different solar indices agree within their
fitting uncertainty, where: $\bar{\gamma_f} = 1.29 \pm 0.07$
and {\color{black}$\bar{\gamma_p} = 3.60 \pm 0.01$}.  %and $\bar{\gamma_p} = 3.61 \pm 0.02$.
The $f$-mode exponent is less than half of the $p$-mode value.
The fundamental mode of solar oscillations has
essentially the character of surface gravity waves
and, contrary to the $p$ modes, it
is essentially incompressible and independent of the hydrostatic structure of the Sun.
Hence, it is not a surprise 
that these different types of modes have different exponents.
Due to their different properties, it is also very likely that different physical 
effects are responsible for their frequency variation.
Accondingly to \inlinecite{DG05}, for the $f$ modes, the dominant cause of
frequency shift is the variation of the subphotospheric magnetic field.
For the $p$ modes, it is
the decrease in the radial component of the turbulent velocity in the outer layers
during the increase in solar activity,
which is accompanied by a decrease in temperature (due to 
a decrease in the efficiency of convective transport).
At low frequency ($\nu < 2.3$ mHz), the $p$-mode frequency shifts have a different behavior
than at high-$\nu$:
a step (with the same exponent $\gamma_p$) or, as found by other authors, 
an exponent twice as large
as the one at high-$\nu$. 
Low-frequency modes have a large probability of being uncorrelated with the solar cycle,
which was not taken into account in the case where a large exponent was estimated.
{\color{black}
The $f$-mode frequency shifts also have a different behavior around 1.7 mHz: they
increase abruptly by an order of magnitude.
}

Modes with frequency around 3 mHz have the smallest probability that their variation
is linearly uncorrelated with the solar index, 
while modes with $\nu < 2.5$ mHz or $\nu > 4.5$ mHz have the largest probability
of being uncorrelated. 
A large  probability ($P_u$) of being uncorrelated does not necessarily means that a given
mode is not physically correlated with the solar cycle, instead it could be due to 
uncertainties in the measurements, a low signal-to-noise ratio. 
The logarithm of $P_u$ is well correlated with
the logarithm of the relative uncertainty of the estimated frequency shift $\delta\nu^e$,
the Pearson correlation coefficient is 
0.71 for medium-$\ell$ modes and 0.54 for high-$\ell$ modes. 
If a given mode has a large probability of being linearly uncorrelated, it is 
expected that the linear fitting of its frequency shifts will have a large uncertainty, hence the
high correlation coefficient between $P_u$ and the estimated frequency shift uncertainty.
However, it raises the question of what could be the physical process 
that would make those modes less sensitive to solar activity.
For a given $\ell$, the upper reflection point for lower-frequency modes is deeper
in the Sun than for high-frequency modes.
If the perturbation layer causing the frequency shift
is above the upper turning point of the mode, 
it would not be affected by the solar cycle.
Accordingly to model S of Christensen-Dalsgaard, 
the depth of the upper turning point increases sharply with decreasing
frequency below 2.3 mHz.
The upper turning point for a radial mode with $\nu = 2$ mHz is
0.5 Mm deeper in the Sun than a three-millihertz mode
(from Figure~2 in \opencite{Cha01}).
At high-frequency, the observed frequency shift seems to suddendly drop to zero.
However, there is no agreement on the exact frequency that this happens,
the observed values range from 3.7 to 5 mHz.
The frequency-shift falloff is explained by an increase of 
chromospheric temperature and magnetic field at solar maximum
(\opencite{Jain96} and references within).
In the presence of an inclined magnetic field,
high-frequency modes tunnel through the temperature minimum
and are particularly sensitive to changes in the chromosphere,
which are expected to be well correlated with solar activity.

%----------------------------------------
\begin{acks}
We are grateful to Tim Larson of Stanford University for 
discussing with us
the results of his improved analysis of MDI medium-$\ell$ data.
The Solar Oscillations Investigation (SOI) involving MDI is supported by NASA
grant NNG05GH14G at Stanford University.  SOHO is a mission of international
cooperation between ESA and NASA. SGK is supported by NASA grant NNG05GD58G.
NOAA Mg {\sc ii} Core-to-wing ratio data are provided by Dr. R. Viereck, NOAA Space
Environment Center.
The solar radio 10.7 cm daily flux (2800 MHz) have been made by the National
Research Council of Canada at the Dominion Radio Astrophysical Observatory,
British Columbia.
The International Sunspot Number was provided by SIDC, RWC Belgium, World Data
Center for the Sunspot Index, Royal Observatory of Belgium.
This study includes data from the synoptic program at the 150-Foot Solar Tower
of the Mt. Wilson Observatory.  The Mt. Wilson 150-Foot Solar Tower is
operated by UCLA, with funding from NASA, ONR, and NSF, under agreement with
the Mt. Wilson Institute.
This work utilizes data obtained by the Global Oscillation Network Group (GONG) program, managed by the National Solar Observatory, which is operated by AURA, Inc.  under a cooperative agreement with the National Science Foundation.  The data were acquired by instruments operated by the Big Bear Solar Observatory, High Altitude Observatory, Learmonth Solar Observatory, Udaipur Solar Observatory, Instituto de Astrof\'isica de Canarias, and Cerro Tololo Interamerican Observatory.

\end{acks}

%%% BIBLIOGRAPHY %%%%%%%%%%%%%%%%%%%%%%%%%%%%%%%%%%%%%%%%%%%%%%%%%%%%%%%%%%%
   
     % format of references provided by the journal (.bst)
\bibliographystyle{spr-mp-sola}

%     % name your Bibtex file containing your references (.bib)
%\bibliography{highl_bibliography_example}  
%
%     % Checking: look if the file containing the ``\bibitem'' exits
%     %           so check if the .bbl file exist (bibTeX compilation)
%\IfFileExists{\jobname.bbl}{} {\typeout{}
%\typeout{****************************************************}
%\typeout{****************************************************}
%\typeout{** Please run "bibtex \jobname" to obtain} \typeout{**
%the bibliography and then re-run LaTeX} \typeout{** twice to fix
%the references !}
%\typeout{****************************************************}
%\typeout{****************************************************}
%\typeout{}}

\end{article} 
\end{document}